\documentclass[graybox]{svmult}

\def\ltsima{$\; \buildrel < \over \sim \;$}
\def\simlt{\lower.5ex\hbox{\ltsima}}
\def\gtsima{$\; \buildrel > \over \sim \;$}
\def\simgt{\lower.5ex\hbox{\gtsima}}
\def\gsimeq
{\hbox{\raise0.5ex\hbox{$>\lower1.06ex\hbox{$\kern-1.07em{\sim}$}$}}}
\def\lsimeq
{\hbox{\raise0.5ex\hbox{$<\lower1.06ex\hbox{$\kern-1.07em{\sim}$}$}}}
\def\pn{\par\noindent}
\def\ss{\smallskip\pn}

\def\xmm{{\it XMM-Newton }}
\def\rosat{{\it ROSAT}}
\def\asca{{\it ASCA}}

\def\xmm{{\it XMM-Newton}}
\def\chandra{{\it Chandra}}
\def\suzaku{{\it Suzaku}}
\def\swift{{\it Swift}}
\def\rxte{{\it RXTE}}
\def\integral{{\it INTEGRAL}}
\def\fermi{{\it Fermi}}
\def\granat{{\it GRANAT}}
\def\herschel{{\it Herschel}}
\def\wmap{{\it WMAP}}
\def\planck{{\it Planck}}
\def\nustar{{\it NuSTAR}}
\def\astroh{{\it Astro-H}}

\def\pg1404{PG~1404+226}

\def\mnras{MNRAS}
\def\apj{ApJ}
\def\aj{AJ}
\def\apjl{ApJL}
\def\apjs{ApJS}
\def\aap{A\&A}
\def\araa{ARA\&A}
\def\nat{Nature}
\def\pasj{PASJ}
\def\prc{Physical Review C}
\def\prd{Physical Review D}
\def\aaps{Astronomy and Astrophysics, Supplement}
\def\pasp{Publications of the ASP}

\def\Fevc{Fe {\sc xxv}}
\def\Fevs{Fe {\sc xxvi}}

\usepackage{mathptmx}       % selects Times Roman as basic font
\usepackage{helvet}         % selects Helvetica as sans-serif font
\usepackage{courier}        % selects Courier as typewriter font
\usepackage{type1cm}        % activate if the above 3 fonts are
\usepackage{makeidx}         % allows index generation
\usepackage{graphicx}        % standard LaTeX graphics tool
\usepackage{multicol}        % used for the two-column index
\usepackage[bottom]{footmisc}% places footnotes at page bottom

\makeindex             % used for the subject index
\begin{document}

\title*{Traces of past activity in the Galactic Centre}
\titlerunning{Traces of past activity in the Galactic Centre}
\author{Gabriele Ponti, Mark R. Morris, Regis Terrier and Andrea Goldwurm}
\institute{Gabriele Ponti \at Max-Planck-Institut f\"{u}r extraterrestrische Physik, 
Giessenbachstrasse 1, D-85748, Garching bei M\"{u}nchen, Germany, \email{ponti@mpe.mpg.de}
\and Mark Morris \at Department of Physics and Astronomy, 
University of California, Los Angeles 90095-1547, USA
\and Regis Terrier \at Unit\'e mixte de recherche Astroparticule et 
Cosmologie, 10 rue Alice Domon et L\'eonie Duquet, 75205 Paris,
France
\and Andrea Goldwurm \at Service d'Astrophysique (SAp), IRFU/DSM/CEA-Saclay,
91191 Gif-sur-Yvette Cedex, France
\at Unit\'e mixte de recherche Astroparticule et 
Cosmologie, 10 rue Alice Domon et L\'eonie Duquet, 75205 Paris,
France
}
\authorrunning{Ponti et al.} 

\maketitle

\abstract{The Milky Way centre hosts a supermassive Black Hole (BH)
with a mass of $\sim4\times10^6$ M$_{\odot}$. Sgr A*, its electromagnetic counterpart, 
currently appears as an extremely weak source with a luminosity $L\sim10^{-9}$ 
L$_{\rm Edd}$. The lowest known Eddington ratio BH. 
However, it was not always so; traces of "glorious" active periods can be found 
in the surrounding medium. 
We review here our current view of the X-ray emission from the Galactic Center (GC) 
and its environment, and the expected signatures (e.g. X-ray reflection) of a past flare.
We discuss the history of Sgr A*'s past activity and its impact on the 
surrounding medium. The structure of the Central Molecular Zone (CMZ) has not 
changed significantly since the last active phase of Sgr A*. This relic torus 
provides us with the opportunity to image the structure of an AGN torus in  
exquisite detail.}

\section{The Galactic Centre (GC)}

At approximately 8 kpc (Reid et al. 2009a,b; Matsunaga et al. 2009; Dambis
et al. 2009; Gillessen et al. 2009; Vanhollebeke et al. 2009; Trippe et al. 2008;
Ghez et al. 2008; Groenewegen et al. 2008; Nishiyama et al. 2006; Bica et al. 2006;
see Genzel et al. 2010 for a review), 
the GC is by far the closest Galactic nucleus (e.g. Andromeda, the nearest 
spiral galaxy, it is $10^2$ times further away). Thus it is an excellent laboratory 
to directly image, with exquisite resolution, the physical processes that might 
be occurring in many other galactic nuclei. 

In the optical window our view toward the Galactic Centre (GC) is obscured 
by a thick dust curtain, with more than 30 magnitudes of extinction (Becklin \& 
Neugebauer 1968; Becklin et al. 1978; Rieke et al. 1989), thus 
preventing any direct view of it. This dust curtain can be raised by 
observations in different wavebands; a direct view of the GC can 
be achieved both at longer and shorter wave-lengths, e.g. in infrared-radio 
as well as medium-hard ($E\geq2-3$ keV) X-rays and $\gamma$-rays. 
InfraRed (IR) and radio observations have demonstrated that the GC is 
very bright and display a stupendous variety of phenomena. 
On large scales ($\sim10$ degrees), a large concentration of stars (the so-called 
Galactic bulge) with a central bar distribution dominates the gravitational 
potential (de Vaucouleurs 1964; Zhao et al. 1994; Blum 1995). 
The central few hundred parsecs contain several of the most massive 
molecular clouds of the Galaxy (Morris \& Serabyn 1996) and radio observations 
reveal arcs and non thermal filaments (Yusef-Zadeh et al. 1984; Morris 1990;
1992; 1996), indicating the importance of the Galactic magnetic field for the GC 
environment (Morris \& Serabyn 1996). 
Within the central few parsecs, IR observations have revealed 
the presence of a dense and luminous star cluster (Genzel et al. 1996; 2000; 
2003; Ghez et al. 2005; 2008; Schodel et al. 2007; 2009; Trippe et al. 2008; 
Lu et al. 2009). 
The precise monitoring of stellar motions within 0.5 arcsec of Sgr A* 
has led to the conclusion that it consists of a compact, dark, central 
mass concentration of $4\times10^6$ M$_\odot$, that cannot be anything 
but a Black Hole (BH, Schodel et al. 2002; Eisenhauer et al. 2003; 
Ghez et al. 2003; 2005; Gillessen et al. 2009; Genzel et al. 2010). 

\subsection{X-ray emission from the GC}
\label{X-rayemission}

The first detections of X-ray sources in the GC date back to the sixties, 
to the very beginning of X-ray astronomy, when the first short (few minute) 
observations were made possible with the use of rocket flights (Bowyer et al. 1965; 
Clark et al. 1965; Fisher et al. 1966; Gursky et al. 1967; Bradt et al. 1968). 
In the 70s the 
first soft X-ray satellites (Uhuru, Ariel 5 and HEAO 1) were launched and by the 
end of the 70s the first GC soft X-ray images, with arcminute resolution, were 
obtained thanks to the Einstein (HEAO 2) satellite (Watson et al. 1981). 
Soon after, in the 80s and early 90s, coded-mask imaging techniques allowed a 
similar revolution in the hard X-ray/soft-$\gamma$-ray spectral regions with 
the launch of 
XRT/SpaceLab2 (Skinner et al. 1987), ART-P/GRANAT (Sunyaev et al. 1991;
Pavlinsky et al. 1994) and SIGMA/GRANAT (Goldwurm et al. 1994; 2001). 
These satellites clearly showed that the soft and hard X-ray emission from the 
GC region is dominated by bright, variable and transient point sources, mainly 
X-ray binaries. 

Koyama et al. (1989), thanks to GINGA data, were the first to discover 
(in the GC) a 6.7 keV line (\Fevc), manifested as diffuse and rather uniform 
emission, resembling the diffuse Galactic ridge emission (Warwick et al. 1985). 
Sunyaev et al. (1993), using ART-P data, discovered an additional harder 
and asymmetric component of the diffuse GC emission. The spatial correlation 
of this emission with GC Molecular Clouds (MC) prompted the speculation 
that MC might act as mirrors and this radiation might be reflection of a past 
flare of a very bright source in the GC (possibly Sgr A* itself). 
In that case, the authors predicted that the Fe K$\alpha$ line, associated with 
the reflection continuum, should soon be detected. 
Three years later, owing to the high energy resolution of the \asca\ satellite, 
the hypothesis of a major X-ray flare in the GC was significantly strengthened 
by the detection of diffuse Fe K$\alpha$ line emission from the hard X-ray 
emitting MC (Koyama et al. 1996). 
Finally, ASCA's improved energy resolution in the soft X-ray band allowed 
also the detection of several high-ionisation emission lines (e.g. of Si, S, Ar, Ca, 
etc.) associated with a low temperature (T$\sim0.8-1$ keV) ionised 
plasma component pervading the GC. 

Since the beginning of this century a new golden age for X-ray astronomy 
has begun, with the launch of XMM-Newton (1999), Chandra (1999) and Suzaku 
(2005), three cornerstones in the ESA, NASA and JAXA space projects 
which have allowed us to take a major step forward in our understanding 
of the GC X-ray emission. 

X-ray observations of the GC show clear evidence for several emission 
components:

{\bf Bright X-ray point sources: }
For more than 3 decades it has been possible to image the GC X-ray 
emission (Watson et al. 1981; Churazov et al. 1994; Pavlinsky et al. 1994;
Sidoli et al. 2001; Sakano et al. 2002; Sidoli et al. 1999; in't Zand et al. 2004). 
In particular, during the past decade, intensive \chandra, \xmm, \rxte, \integral\ 
and, on an almost daily basis, \swift\ monitoring campaigns have been performed
within the central 1.2 square degrees (Wang et al. 2002; Baganoff et al. 2003;
Muno et al. 2003; 2004; 2006; 2009; Kennea et al. 2006; Degenaar et al. 2009; 
2010; 2012; Swank et al. 2001; Kuulkers et al. 2007). 
Nineteen sources having a 2-10 keV peak luminosity, L$_{\rm x} \gsimeq 10^{34}$ 
erg s$^{-1}$,
have been detected in this region so far (Degenaar et al. 2012; Wijnands et al. 2006). 
It has been possible to identify the brightest ones (about half of the sample) as 
accreting neutron stars or BH.
Two of the nineteen sources are persistent (with L$_{\rm x} \simeq 10^{35-36}$ 
erg s$^{-1}$), while the remaining are transients spending years to decades in 
a quiescent state with L$_{\rm x} \lsimeq 10^{33}$ erg s$^{-1}$ and undergoing 
short (weeks/months) outbursts with peak luminosity several orders of magnitudes 
higher. Ten of the 19 bright sources in the central 1.2 square degrees belong 
to the class of low-luminosity X-ray transients and have never been observed 
at luminosity higher 
than L$_{\rm x} \sim 10^{36}$ erg s$^{-1}$ and only 3 have reached L$_{\rm x} 
\gsimeq 10^{37}$ erg s$^{-1}$ (Degenaard et al. 2012; Wijnands et al. 2006). 
Moreover, no new transients have been found since 2007, suggesting that 
most X-ray transients recurring on a time scale of less than a decade have 
now been identified near the GC (in 't Zand et al. 2004; Muno et al. 2009; 
Degenaar et al. 2010; 2012).

{\bf Soft X-ray diffuse plasma: }
A diffuse soft X-ray emission (E$_{\rm x} \lsimeq 4$ keV), traced by the H-like and 
He-like lines of, e.g., Si, S, Ar \& Ca can be best fitted with a low temperature 
(kT$\sim0.8-1$ keV) plasma component that pervades the GC (a small fraction, 
less than a few percent, of this emission can be ascribed to foreground diffuse 
soft X-ray emission; Kuntz \& Snowden 2008; Yoshino et al. 2009; Masui 
et al. 2009; Kimura 2010). 
This plasma is bound to the Galaxy. Muno et al. (2004), in fact, estimated the 
sound speed for the soft plasma to be $v_s\sim500$ km s$^{-1}$. This 
plasma will possibly expand to $\sim1$ kpc above the disc, but be 
bound to the Galaxy (assuming the Breitschwerdt et al. 1991 potential). 
The plasma radiative cooling time, within the central 50 pc, is estimated to 
be $3-6\times10^6$ yr (Muno et al. 2004). If about 1 \% of the $10^{51}$ 
erg of kinetic energy per supernova heats the plasma (as in the Galactic 
disc; Kaneda et al. 1997), it can be sustained by the reasonable rate of 
one supernova every $10^5$ yr. 
Additional soft X-ray emission might be produced by interaction of 
stellar winds from young and massive stars with the interstellar medium 
(Muno et al. 2004) or each other (Mauerhan et al. 2009) and coronal 
X-ray sources (Favata et al. 1997; Covino et al. 2000). 
The patchy soft plasma distribution, with higher concentrations (up to a 
factor of 10 variations within 50 pc) towards star forming regions, 
matches these hypotheses. 

{\bf Diffuse 6.7 keV (\Fevc) emission and Galactic ridge X-ray emission: }
Discovered in the late 70's (Cooke et al. 1970; Bleach et al. 1972;
Worral et al. 1982), the Galactic ridge X-ray emission is one of the 
largest diffuse features (after the X-ray background) in the 2-10 keV band. 
With a total X-ray luminosity of L$_{\rm x} = 1-2\times 
10^{38}$ erg s$^{-1}$ (Yamauchi et al. 1993; Valinia et al. 1998; Revnivtsev 
et al. 2006), it is concentrated near the Galactic disc and bulge, extending 
more than 100 degrees in Galactic longitude and just a few degrees 
in latitude (Warwick et al. 1985; 1988; Yamauchi et al. 1990; 1993; 
Revnivtsev 2003). With strong emission lines (e.g. \Fevc\ and \Fevs\ 
at 6.7 and 6.97 keV) and a hard X-ray continuum, this emission 
has a spectrum characteristic of a kT$\sim5-10$ keV, optically thin plasma 
(Koyama et al. 1986; 1989; 2007; Yamauchi et al. 1993; 1996; 2008).
However, such a hot plasma cannot be gravitationally or magnetically bound 
to the Galaxy (Tanaka 2002), thus it would flow away at a supersonic velocity 
of few thousands km s$^{-1}$ (Zel'dovich \& Rayzner 1966), on a timescale 
of $\sim3\times10^4$ years (Sunyaev et al. 1993). To replenish the energy 
losses a luminosity (supplied throughout the Galaxy) of $10^{43}$ erg s$^{-1}$ 
would be required (Tanaka 2002). At present, no evidence exists for the presence 
of such a source in the Galaxy.

Alternatively, this diffuse emission might be composed of a large population of 
initially unresolved, weak point sources (Revnivtsev et al. 2009), as is the 
extragalactic X-ray background. 
In support of this hypothesis, it is observed that both the \Fevc\ 6.7 keV line 
and the hard X-ray (3-20 keV) continuum are very well correlated, over the whole 
Galaxy, with the near IR luminosity ($3-4 \mu$m) which traces the stellar mass 
density (Revnivtsev et al. 2006a,b). Moreover, thanks to an ultra-deep (1 Ms) 
\chandra\ observation of a field on the Galactic plane (l=0.08, b=-1.42, near the GC) 
with weak interstellar absorption, Revnivtsev et al. (2009) resolved more than 
80 \% of the diffuse 6-8 keV emission into weak discrete sources such as accreting 
white dwarfs and coronally active stars. Additionally, the observed luminosity function 
of these sources is observed to be very similar to the one present in the Solar 
vicinity (Sazonov et al. 2006; Revnivtsev et al. 2007; 2009). 
The still unresolved 10-20 \% diffuse emission is consistent with being 
composed of coronally active and normal stars with L$_{\rm 2-10 keV} < 4 
\times 10^{29}$ erg s$^{-1}$, thus below the detection threshold (Revnivtsev 2009). 
It now seems clear that the Galactic plane 6.7 keV emission (the so called 
Galactic ridge diffuse X-ray emission) is mainly composed of weak point sources.

Strong 6.7 keV emission is also present in the GC (Koyama et al. 1989; Yamauchi 
et al. 1990; Nottingham et al. 1993). As for the Galactic ridge, part of 
this emission will be produced by accreting white dwarfs and coronally active stars.
However, a significant enhancement of 6.7 keV emission, as well as an 
east-west asymmetry compared to the near IR distributions (tracing point sources), 
are observed in the central degree (Koyama et al. 2007; 2009; 2011; Yamauchi et al. 
2009; Uchiyama et al. 2011). 
This GC excess of emission (over the best-fit stellar mass distribution model 
made from near-IR observations and explaining the 6.7 keV emission in the Galactic 
ridge) is hardly explained by point source origin, requiring a new population of 
sources with extremely strong \Fevc\ line emission (Uchiyama et al. 2011; 
Tanaka \& Yamauchi 2011).
Thus, it has been suggested that this component might be associated with truly 
diffuse hot gas (Koyama et al. 2007; Uchiyama et al. 2011). 
The energy required to sustain such 
hot plasma within the inner 20 pc of the Galaxy is 3 orders of magnitude smaller 
than for the entire Galaxy, being $\sim10^{40}$ erg s$^{-1}$ (Muno et al. 2004). 
Belmont et al. (2005) explored the possibility of the hot gas being a 
gravitationally confined helium plasma. Belmont \& Tagger (2006), instead, 
suggested that viscous friction of MC flowing within the hot phase can dissipate 
energy in the gas and heat it. 
As for the Galactic plane, \chandra\ can be used to partially resolve the point 
source contribution to this GC excess of emission. 
The lower extinction at a distance of 4-9 pc from Sgr A*, allowed Revnivtsev 
et al. (2007) and Yuasa et al. (2012) to resolve at least 40 \% and possibly 
the bulk of the total 4-8 keV band GC emission into (as in the Galactic ridge) accreting white dwarfs and coronally active stars. 
The remaining GC 6.7 keV emission might be associated with supernova remnants, 
such as Sgr A East (Maeda et al. 2002; Sakano et al. 2004; Koyama et al. 2007) 
which not only show a low temperature (kT$\sim0.8-1$ keV) plasma component, 
but also a significantly hotter (kT$\sim3-4$ keV) component producing a harder 
X-ray continuum and significant \Fevc\ line emission. 

However, it is difficult to completely rule out that some additional emission might 
be associated with a hot plasma produced by past energetic events in the GC. 

\label{SgrA*}
{\bf Sgr A*: }
One of the major drivers for the deep X-ray surveys of the GC during the 
past half a century, has been to measure the X/$\gamma$-ray emission 
from the supermassive BH associated with the compact radio source Sgr A* 
(Melia \& Falcke 2001; Melia 2007).
A weak source has been detected, at the position of Sgr A*, with a 2-10 keV 
luminosity of only L$_{\rm 2-10}=2\times10^{33}$ erg s$^{-1}$ 
(Baganoff et al. 2003), which corresponds to $4\times10^{-12}$ of the Eddington 
luminosity for a $4\times10^6$ M$_\odot$ BH. No other BH has ever been detected 
at such low Eddington ratio. The excellent \chandra\ spatial resolution allowed also to 
beautifully resolve the extent of Sgr A*'s quiescent emission to slightly more than 1" 
(Baganoff et al. 2003), which roughly corresponds to the Bondi radius (Bondi 1952; 
$r_{\rm Bondi}\sim2\times10^5$ r$_{\rm g}$, where r$_{\rm g}=GM_{\rm BH}/c^2$) and to 
detect diffuse emission (within 15" radius from Sgr A*) from a 
non-ionisation-equilibrium plasma (Xu et al. 2006). 
Another major result has been the discovery of Sgr A*'s flaring behaviour, both 
in X-rays (Baganoff et al. 2001; Goldwurm et al. 2003) and near-infrared (Genzel 
et al. 2003; Ghez et al. 2004). X-ray flares typically occur about once per day, 
lasting for $\sim10^{3-4}$ s, during which the X-ray intensity increases by factors up 
to 160 (Porquet et al. 2003) from the quiescent value, however the bolometric 
luminosity still remains extremely low. 
All these observational properties appear to be consistent with Sgr A* 
accreting stellar wind material captured from nearby stars 
(Melia et al. 1992; Coker \& Melia 1997; Rockefeller et al. 2004; 
Quataert 2004; Cuadra et al. 2005; 2006; 2008).

In the past decade Sgr A* has been repeatedly observed by \xmm\ and \chandra\ 
(as well as in IR), accumulating more than 3 Ms exposure (corresponding to 
almost two months) and confirming the Sgr A*'s low quiescent-state luminosity. 
One may wonder whether Sgr A* has always been so underluminous or whether 
it experienced, in the past, long periods of high-energy activity, that would make 
the massive black hole of our Galaxy more similar to typical low-luminosity 
Active Galactic Nuclei, than it appears today.

{\bf Diffuse 6.4 keV (Fe K$\alpha$) emission: } 
Extended ($\sim1^{\circ}$, few $10^2$ pc), hard (8-22 keV) X-ray emission 
elongated along the Galactic plane, corresponding spatially with massive 
Molecular Clouds (MC) was first detected in the early 90's thanks to the 
ART-P telescope (Sunyaev et al. 1993). The authors interpreted this emission 
as X-ray reflection of a past energetic event in the GC, predicting the presence 
of the associated fluorescent Fe K$\alpha$ line\footnote{Even before 
the Sgr A* case, the (delayed) reflected emission has been discussed 
(see Sunyaev et al. 1991; Churazov et al. 1993) for 1E1740.7-2942, one 
of the two brightest hard X-ray sources in the GC region. 
During the mid 90s the source, known to be located inside a molecular 
cloud (Bally \& Leventhal 1991), was observed in a low state with the 
flux $\sim5$ times lower than in the bright state. In a set up similar 
to the one used later for Sgr A*, this fact was used to 
constrain the optical depth and the size of the cloud from 
hard X-rays and later from 6.4 keV line (Churazov et al. 1996).
}. 
Intense, diffuse and patchy Fe K$\alpha$ emission, 
from the same MC, was later measured by ASCA (Koyama et al. 
1996), confirming this interpretation. 
The spatial and temporal variations of the scattered/reflected flux depend on the 
underlying gas distribution and the past activity of the GC source over a time equal 
to that taken by the light to cross that region. The presence of massive MC in the 
inner core of the Galaxy (the so called Central Molecular Zone, CMZ) allows us 
to put firm constraints on the past activity of the GC over the last $\sim10^3$ yr.
This remains true even if other mechanisms can explain the diffuse 
Fe K$\alpha$ emission and hard continuum (Valinia et al. 2000; Yusef-Zadeh et al. 
2002; 2007; Bykov 2003; Dogiel et al. 2009; 2011). 

\section{X-ray reflection}
\label{refl}

\begin{figure}[t]

\hspace{-2.0cm}
\includegraphics[width=0.63\textwidth,height=0.7\textwidth,angle=-90]{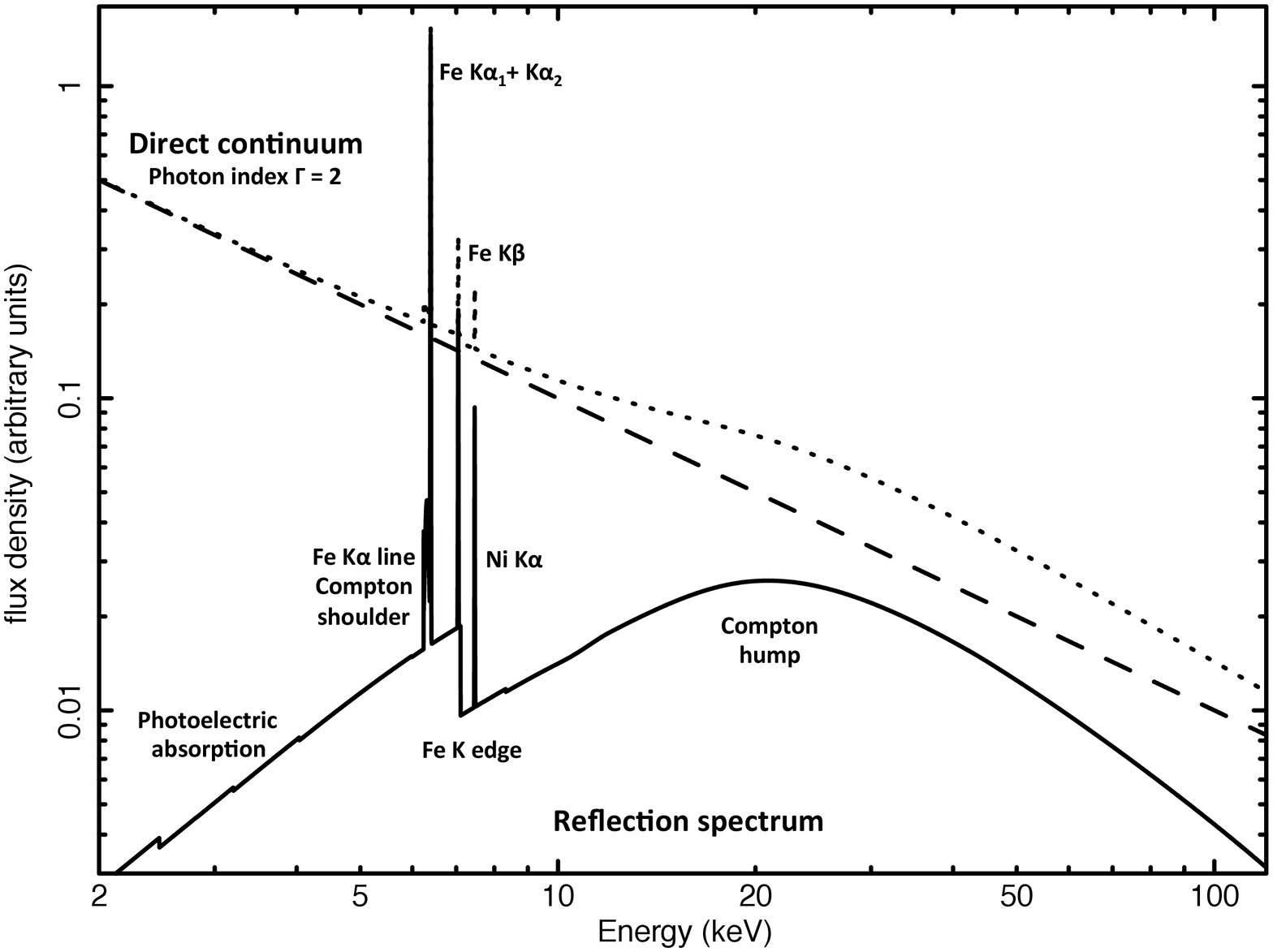}

\vspace{-7.85cm}

\hspace{+4.8cm}
\includegraphics[width=0.76\textwidth,height=0.715\textwidth,angle=0]{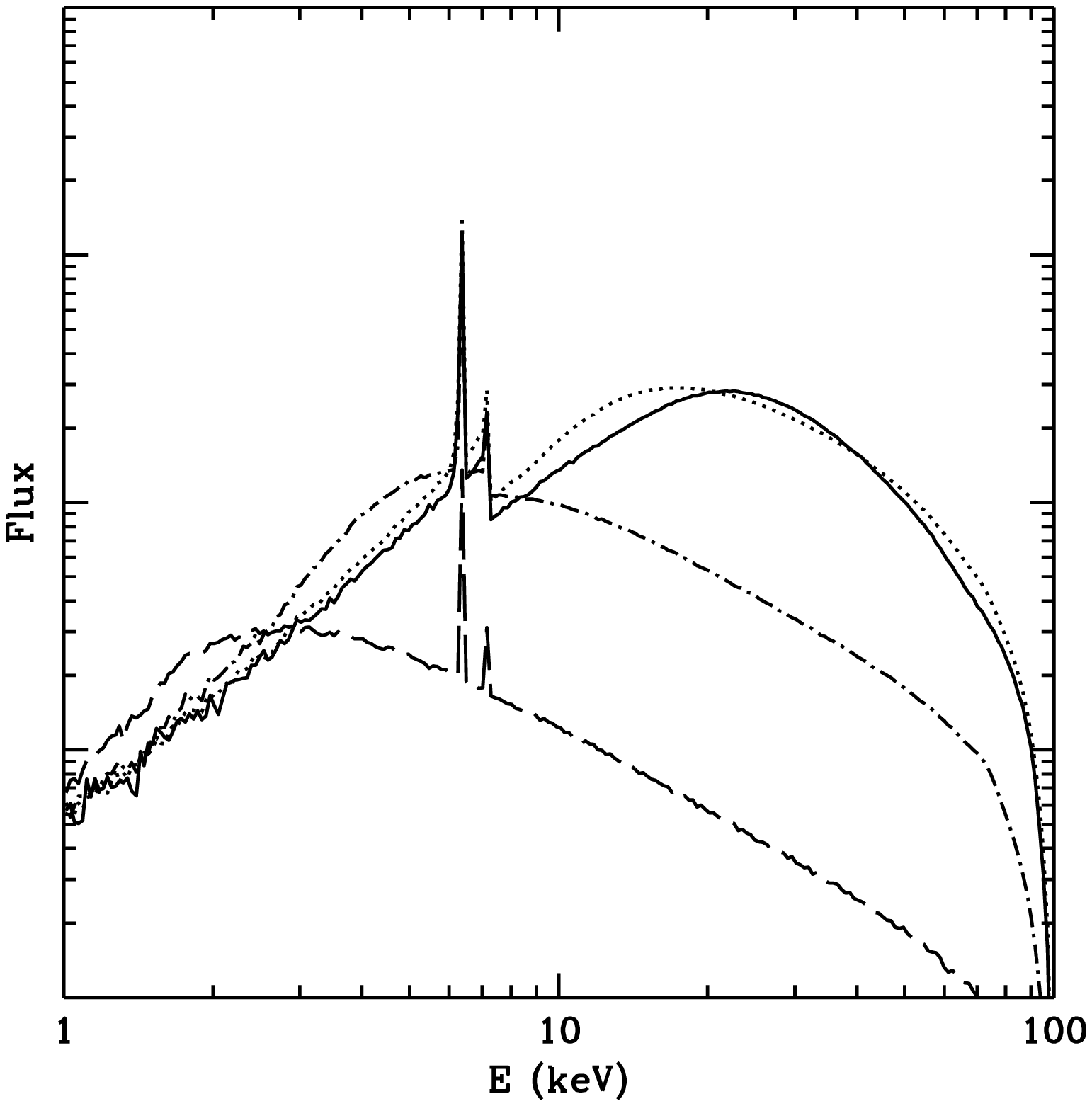}
\caption{{\it (Left panel)} Typical X-ray reflection spectrum from neutral material
with column densities $\sigma_T\gg1$. 
The Fe K$\alpha$ doublet + K$\beta$ and Ni K$\alpha$ lines and the 
respective Compton shoulders, as well as the K edges and Compton hump
are shown (emission lines from lower atomic number elements which have 
lower EW are not shown). 
{\it (Right panel)} Reflection spectrum from a molecular cloud 
(with toroidal geometry and observed face on) for different column 
densities: $2\times10^{22}$ cm$^{-2}$ (dashed line), 
$2\times10^{23}$ cm$^{-2}$ (dot-dashed line), $2\times10^{24}$ cm$^{-2}$ (dotted line), 
$2\times10^{25}$ cm$^{-2}$ (solid line). A prominent Compton hump is expected for 
$\tau\gg1$, while the high-energy spectrum changes significantly for smaller 
column density clouds. Almost no Compton hump is visible for $N_H<2\times10^{23}$
cm$^{-2}$. [Figure courtesy of Matt et al. 2003]
}
\label{Refl} 
\end{figure}
The radiation produced by a powerful X-ray source located close to a MC will 
interact with the MC material and generate an X-ray reflection component. 
Hard X-ray radiation entering a MC (with typical hydrogen column density 
of the order of $N_{\rm H} \sim$ few $10^{22} - 10^{24}$ cm$^{-2}$; 
i.e. $\tau \sim 0.01 - 1$) produces, in fact, an X-ray reflection spectrum similar 
to the one shown in Fig. \ref{Refl}. We describe here the main features 
of the X-ray reflection from (neutral) MC material. The illuminating source 
is assumed to have a power law spectrum with photon index $\Gamma=2$. 

\label{Chump}
{\bf Continuum: } Low energy X-ray radiation is mainly absorbed 
through the photoionisation of hydrogen, helium and the K-electrons of heavy 
elements. The photoionisation cross section rapidly decreases with frequency: 
$\sigma_{\rm ph}\sim\nu^{-3}$, thus the Thomson scattering cross section 
($\sigma_{\rm T}=6.65\times10^{-25}$ cm$^{-2}$) already exceeds the 
photoionisation cross-section per hydrogen atom for photons with energies 
E$>8-12$ keV (Bethe \& Salpeter 1957; Brown \& Gould 1970; Basko et al. 1974). 
At energies E$>$few keV the photon wavelength is less than the Bohr radius 
and electrons bound to hydrogen and helium atoms can be treated as free 
since the recoil energy $\sim h\nu(h\nu/m_ec^2)$ greatly exceeds the 
ionisation potential of these atoms (Basko et al. 1974). 
High-energy photons lose energy mainly through the recoil effect, 
transferring it to the electrons at each scattering. 
The competition between multiple electron (down-) scattering 
of high-energy photons and photoelectric absorption of low energy photons leads 
to the characteristic "Compton reflection hump" in the reflection spectrum 
between 20-100 keV (Basko et al. 1974; Fabian 1977; Basko 1978; 
George \& Fabian 1991; Matt et al. 1991; Nandra \& George 1994; Ghisellini et al. 1994).
The left panel of Fig. \ref{Refl} shows the reflection spectrum from high optical 
depth ($\tau\gg1$) MC. A prominent Compton hump is present between 10 and 
50 keV. The right panel shows the reflection spectrum from clouds with lower column 
densities. The reflection spectrum has a very different shape and the Compton 
hump is less prominent.

{\bf Edges and lines: } The reflection spectrum does contain jumps corresponding 
to the K-edges of photo-absorption by heavy elements such as Fe, S, Ar, etc. 
Due to the high abundance and fluorescent yield of iron, about 30 \% of the 
photons absorbed by photoionisation of iron with energies above the K-shell 
edge at 7.1 keV give rise to iron K-shell fluorescent photons. 
The Fe K$\alpha$ line (2p$  \to $1s transition) consists of two components 
K$\alpha_{1}$ at 6.404 and K$\alpha_{2}$ at 6.391 keV with branching ratio 
of 2:1 (Bambynek et al. 1972). The lines have natural widths of 
$\Delta E\sim3.5$ eV. High resolution X-ray spectrometers (such as 
the one provided by X-ray calorimeters) will resolve the doublet and 
measure with high accuracy any shift due to the cloud motion. 
For a MC moving at 100 km s$^{-1}$ a 2.1 eV shift is expected. 
Since the M-shells of neutral iron atoms are populated, the 
fluorescent Fe K$\beta$ line 
(3p$\to$1s) at 7.06 keV, with K$\beta$/K$\alpha$ ratio =0.155-0.16 
(Molendi et al. 2003; Basko 1978; Palmeri 2003a,b), is also expected.

Because both the Fe K$\alpha$ and reflection continuum intensities 
are roughly proportional to $\tau_T$ and source luminosity the 
Equivalent Width, EW$_{\rm Fe K\alpha}$ is expected to be $\sim 1$ keV (
for Solar abundances), with only $\sim30$ \% variations (from $\sim0.7$ to 
$\sim1$ keV; see Fig. 3 of Matt et al. 2003) for different MC column densities 
(Matt et al. 2003). 
The line EW depends also on the system geometry and the iron 
abundance (A$_{\rm Fe}$). For Solar abundances (A$_{\rm Fe}=1$) 
the line EW is $\sim1$ keV against the total reflection continuum (Ghisellini 
et al. 1994; Krolik et al. 1994; Matt et al. 1996; 2003) and it is expected 
to vary approximately linearly 
EW$\sim$EW(A$_{\rm Fe}=1$)$\times$A$_{\rm Fe}^{0.8-0.9}$
for A$_{\rm Fe}<<1$ and saturates for A$_{\rm Fe}>>1$, 
EW$\sim$EW(A$_{\rm Fe}=1$)$[1+ (0.6-1.2) \times $log(A$_{\rm Fe}$)] 
(Matt et al. 1997). 
Because of the angular dependence of the Compton effect, the 
intensity of the scattered continuum (for $\tau\ll1$) measured at 
an angle $\theta$ with respect to the direction of travel of the 
incident radiation is expected to be proportional to 
$\propto (1+cos^2(\theta))$. On the contrary Fe K$\alpha$ fluorescence 
emission is isotropic, thus the observed EW$_{\rm Fe}$ has also 
a dependence on the system geometry.  
The K-shell fluorescence yield ($Y^K_Z$) is an increasing 
function of atomic number (Bambynek et al. 1972), thus weak lines are expected 
from C, N, O, and Ne ($Y^K_Z\sim10^{-2}$) as well as Mg, Si and S 
($Y^K_Z\sim10^{-1}$). EW$_{\rm Ni K\alpha}$ is estimated to be 
$\sim 0.05 \times$EW$_{\rm Fe K\alpha}$ (Yaqoob \& Murphy 2011). 

{\bf Compton line shoulders: } The recoil upon electron scattering shifts 
the Fe K$\alpha$ line photons to lower energies (but this is true for any line). 
Thus, the initial mono-energetic line spreads downward in energy over two 
Compton wavelengths ($\Delta E \simeq2E^2/m_e c^2=160$ eV) for each 
scattering, in this way forming a red wing, the so called "Compton shoulder". 
The details of the Compton shoulder EW and shape depend on the geometry,
and the MC column density. For example, for small scattering angles of 
X-ray photons the recoil energy is comparable or lower than the ionisation 
energy (Sunyaev \& Churazov 1996), leading to modifications of the shape 
of the Compton shoulder. The motion of the electrons in hydrogen atoms 
is another important factor causing strong blurring of the low energy wing 
of the shoulder (Sunyaev \& Churazov 1996). However, in most cases 
the Compton shoulder EW is expected to be roughly 10-20 \% of the 
Fe K$\alpha$ EW (Matt 2002; Yaqoob \& Murphy 2010). 

{\bf Polarisation: } Thomson scattering on electrons will induce 
linear polarisation of the reflected emission, even in the case of 
initially unpolarised radiation. The expected degree of polarisation 
is given by $P=(1-cos^2\theta)/(1+cos^2\theta)$, where $\theta$ is the 
scattering angle (Basko et al. 1974; Churazov et al. 2002). 
The detection of polarised light from the MC in the GC will 
provide a clean test of the X-ray external illumination scenario. 
Moreover, the scattering angle deduced from polarisation (Sunyaev \& Churazov 
1998; Odaka et al. 2011) will eventually allow one to determine the 
MC distance along the line of sight. 
This is a fundamental parameter for reconstructing the three dimensional 
gas distributions in the CMZ and the recent history of GC X-ray emission. 

{\bf Dust scattering halo:} The canonical extinction 
towards the GC is $A_V\sim 30$ mag, which translates to an X-ray 
absorbing column density of $N_H\sim6\times10^{22}$ cm$^{-2}$
(Predehl \& Schmitt 1995; Fritz et al. 2011). 
The interaction of X-rays with interstellar 
dust grains leads not only to mere absorption, but also to small-angle 
scattering, producing a faint and diffuse X-ray halo around the source 
(Overbeck 1965; Martin 1970; Toor et al. 1976; Rolf 1983; 
Predehl \& Schmitt 1995; Xiang et al. 2007; 2011). Because of the 
energy dependence of the scattering ($I_{halo}\sim E^{-2}$), dust 
scattering halos usually appear as a soft power-law component 
with a fractional intensity of a few to few tens of per cent. 
The spectral shape of any source in the GC will be 
modified by the presence of the intervening material. 
Detailed studies of the dust-scattering halo of bright sources can, 
in principle, allow one to derive the distribution, along the line of sight, 
of the scattering material. 

\subsection{Time dependent effects}

Depending on the mutual position of the MC and the primary 
source, substantial evolution of the morphology, flux, equivalent 
width and shape of the Fe K$\alpha$ emission and reflection 
continuum is expected as a response to a variable primary 
illuminating source (Sunyaev \& Churazov 1998; Odaka et al. 2011). 
The light crossing time of the CMZ is t$\sim10^{3}$ yr, thus monitoring of the  
reflection component from MC provides constrains on the GC X-ray emission 
on a similar time-scale and also suggests that the observed present-day 
X-ray reflection component might be associated with a presently dim source 
that was brighter in the past (Sunyaev et al. 1993; Koyama et al. 1996). 

To a first approximation it is possible to place a direct link between the illuminating 
source luminosity and the Fe K$\alpha$ reflection line flux, assuming 
isotropic source emission, outside of a cloud with $\tau_T\ll1$ 
(see e.g. Sunyaev \& Churazov 1998):
\begin{equation}
F_{\rm FeK\alpha}= 
\phi 10^7 \frac{\Omega}{4 \pi D^2} \frac{\delta_{Fe}}{3.3\times10^{-5}} \tau L_8 
~~~~~({\rm photon~s^{-1}~cm^{-2}}), 
\label{eq}
\end{equation}
where $\phi$ is a factor of order unity which weakly depends on 
the source spectral shape; $\Omega$ is the solid angle of the cloud subtended 
from the location of the primary source; $D$ is the distance to the observer; 
$\delta_{Fe}$ is the iron abundance with respect to hydrogen; $\tau$ is 
the optical depth of the cloud; $L_8$ is the continuum source luminosity 
(erg s$^{-1}$) at 8 keV in the 8 keV wide energy band (see Sunyaev \& 
Churazov 1998 for definition). Thus: 
$L_8=6\times10^{38}\times (F_{\rm FeK\alpha}/10^{-4}) \times 
(0.1/\tau_T) \times (\delta_{Fe}/3.3\times10^{-5})^{-1} \times (R/100~{\rm pc})^2$ 
(erg s$^{-1}$),
where R is the distance from the source to the cloud. 
If the duration of the source flare is shorter than the cloud light crossing 
time (r$_{\rm cloud}$/c), the luminosity estimates should be multiplied by a factor roughly 
r$_{\rm cloud}$/c$\Delta$t. 
Because both the Fe K$\alpha$ and reflection continuum intensities 
are roughly proportional to $\tau_T$ and source luminosity the EW$_{\rm Fe K\alpha}$ 
is expected to be $\sim 1$ keV (for Solar abundances), with a weak dependence on
the MC parameters (Matt et al. 2003). 

{\bf Light front scanning the CMZ: } The right panel of Fig. \ref{CMZscan} shows the 
situation when the CMZ is illuminated by two very short flares from Sgr A*. 
The light echo seen from a point different from the source is described by an 
ellipse, i.e. the locus of points so that the sum of the distances to the source and 
observer is constant. Being the observer at large distance, one focus goes at 
infinity and the ellipse becomes a parabola. 
The echo will appear as a parabola, with rotational symmetry about the z-axis, 
with equation $z/c=(t^2-(x/c)^2)/2t$, to an observer at infinity (Churazov \& Sunyaev 1998). 

The propagation of an echo in the CMZ allows us to scan the CMZ gas distribution 
and to perform a tomography of the CMZ. Different patterns of variability are 
expected for short and long duration flares. In general, the distinction between 
short and long duration is made on the basis of the light crossing time of the 
MC and/or MC substructure under investigation. A short duration flare (e.g. months) 
is expected to produce strong variations of the surface brightness of the 
Fe K$\alpha$ line across the MC image, on angular scales corresponding 
to the typical size of the MC non-uniformities (down to few arcsec), on the other 
hand a long duration flare (decades) is expected to reflect the total MC optical 
depth, which will be much smoother. Monitoring campaigns of the evolution
of the Fe K$\alpha$ morphology (in connection with MC column density maps) 
will allow a disentangling of these different scenarios. 

{\bf Superluminal propagation of the echo: } The apparent echo expansion 
velocity over the x-coordinate (see Fig. \ref{CMZscan}) is equal to or larger than the 
speed of light for any position ( $|\dot{x}/c|=1+(ct-|x|)^2/2ct|x|$ ). Similarly, 
if the MC is located closer to us than the source, then the echo will scan 
the MC much quicker than the light-crossing time of the cloud 
($\dot{z}/c=0.5+x^2/2(ct)^2$, thus $\dot{z}>c$ if $z<0$, when the MC 
is in front of the source). Superluminal echos can also be observed in 
particular geometrical configurations and are mainly linked to the fact that 
no causal connection is present between the illuminated parts of the MC. 
The framework is very similar to a projector-screen scenario. The 
physical variation happens at the level of the projector, for example 
changing the slide (e.g. taking a time $t_s$). This variation will 
appear projected over a distance $d_{wall}$, thus moving at a velocity 
$v=d_{wall}/t_s$ which can be $v=d_{wall}/t_s\gg c$ as long as the 
projector is powerful enough to illuminate a distant enough MC. 
Observations of superluminal echos have been, and will be, powerful tools 
to establish the origin of the external illumination of the Fe K$\alpha$ emission 
(Ponti et al. 2010). 

{\bf Propagation of the Fe K$\alpha$ morphology: } The morphology and the 
time evolution of the Fe K$\alpha$ emission also provide important 
constraints, when compared to the MC $N_H$ distribution, on the location of the 
illuminating source (Churazov \& Sunyaev 1998; Murakami et al. 2001; Odaka 
et al. 2011). A general trend appears where the MC starts to light up from the 
direction of the source and, if the MC is optically thin ($\tau_T\ll 1$), the surface 
brightness distribution reflects the density distribution in the cloud along the 
surface of the parabola, disappearing on the time-scale of the flare, convolved 
with the light crossing time of the cloud. In thicker clouds the light front 
might not penetrate the denser condensations which will then appear as holes 
(or cast shadows on the more distant parts of the cloud) in the Fe K$\alpha$ 
emissivity if the source is located behind (in front of) the MC 
(Churazov \& Sunyaev 1998; Murakami et al. 2001; Odaka et al. 2011). 
The details of the Fe K$\alpha$ morphology thus primarily depend on the 
MC $N_H$ distribution, the duration and light curve of the flare and the 
source-MC-observer geometry.

{\bf Multiple scatterings: } Even when the light front has already passed through 
the MC, multiply scattered photons can still be 
observed. For example the photons forming Compton shoulders of the lines 
are created by Thomson scattering of the line photons, thus they will leave the 
MC with some delay (typically of the order of the light-crossing time of the MC) 
compared to the primary radiation. 
Additionally, optically thick and dense MC cores, where multiple scatterings occur, 
should also still shine for some time after the light front passage. 
Another effect of multiple scatterings is to dilute the response of the reflection 
component to fast source variability happening on time-scales shorter than 
the typical scattering time. 

\subsection{X-ray induced chemistry and MC heating}

An X-ray light front irradiating a MC is not only expected to produce an X-ray 
reflection component, but also to interact with molecules, thus inducing 
chemical reactions. A correlation between the distribution of Fe K$\alpha$ 
and rotational line emission from the refractory molecule SiO (with an increase 
by a factor of $\gsimeq20$) has been observed by Mart\'in-Pintado et al. (2000). 
This was an indication that either X-rays irradiation or something associated 
with that is removing the silicon from dust grains in the form of SiO.  
Amo-Baladr\'on et al. (2009) confirmed and strengthened this result and 
suggested that the release of SiO from small grains can occur when hard 
X-rays deposit their energy in the grains, and heat them locally to high 
temperature (up to $\sim10^3$ K). 

X-ray irradiation might also be the cause of the high temperature of the 
MC of the CMZ. Rodr\'iguez-Fern\'andez et al. (2004) excluded that X-ray 
irradiation can be an important source of MC heating. However, they 
examined present day X-ray luminosities, whereas much brighter past 
X-ray flares may have occurred at typical intervals much shorter than 
the cooling time of the MC, so they could be responsible for an important 
portion of the heating of MC in the GC.

\section{The GC environment}
\label{GCenvironment}

\begin{figure}[t]
\hspace{-4.5cm}
\includegraphics[scale=0.65,angle=-90]{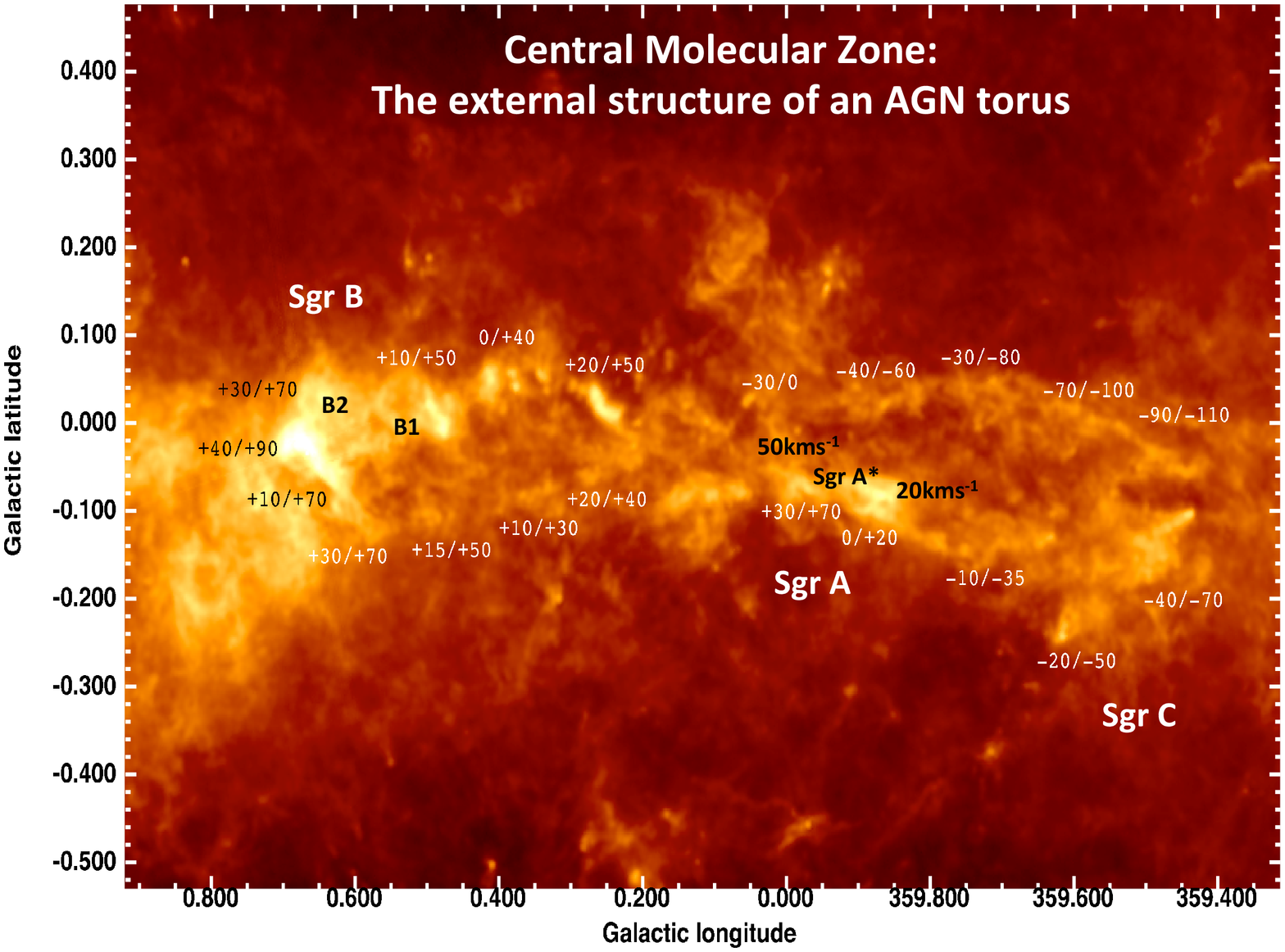}
\caption{\herschel\ map of the atomic hydrogen column density 
distribution of the central $\sim150 pc$ 
of the CMZ showing a large part of the the CMZ disc population.
The color scale is logarithmic and extends from $4\times10^{22}$ 
cm$^{-2}$ in the darkest regions to $4\times10^{25}$ cm$^{-2}$ in the 
brightest MC cores. Velocity information is taken from CS spectroscopic cubes 
(Tsuboi et al. 1999) for the gas counterparts positionally associated with the
dense dust clumps. 
The CMZ disc population appears to have a geometry and location remarkably 
similar to the one of the AGN molecular torus invoked by AGN unification schemes 
(Urry \& Padovani 1995). [Figure courtesy of Molinari et al. 2011]}
\label{CMZ} 
\end{figure}

\subsection{The Central Molecular Zone}

\begin{figure}[t]
\hspace{-0.7cm}
\includegraphics[scale=0.26,angle=0]{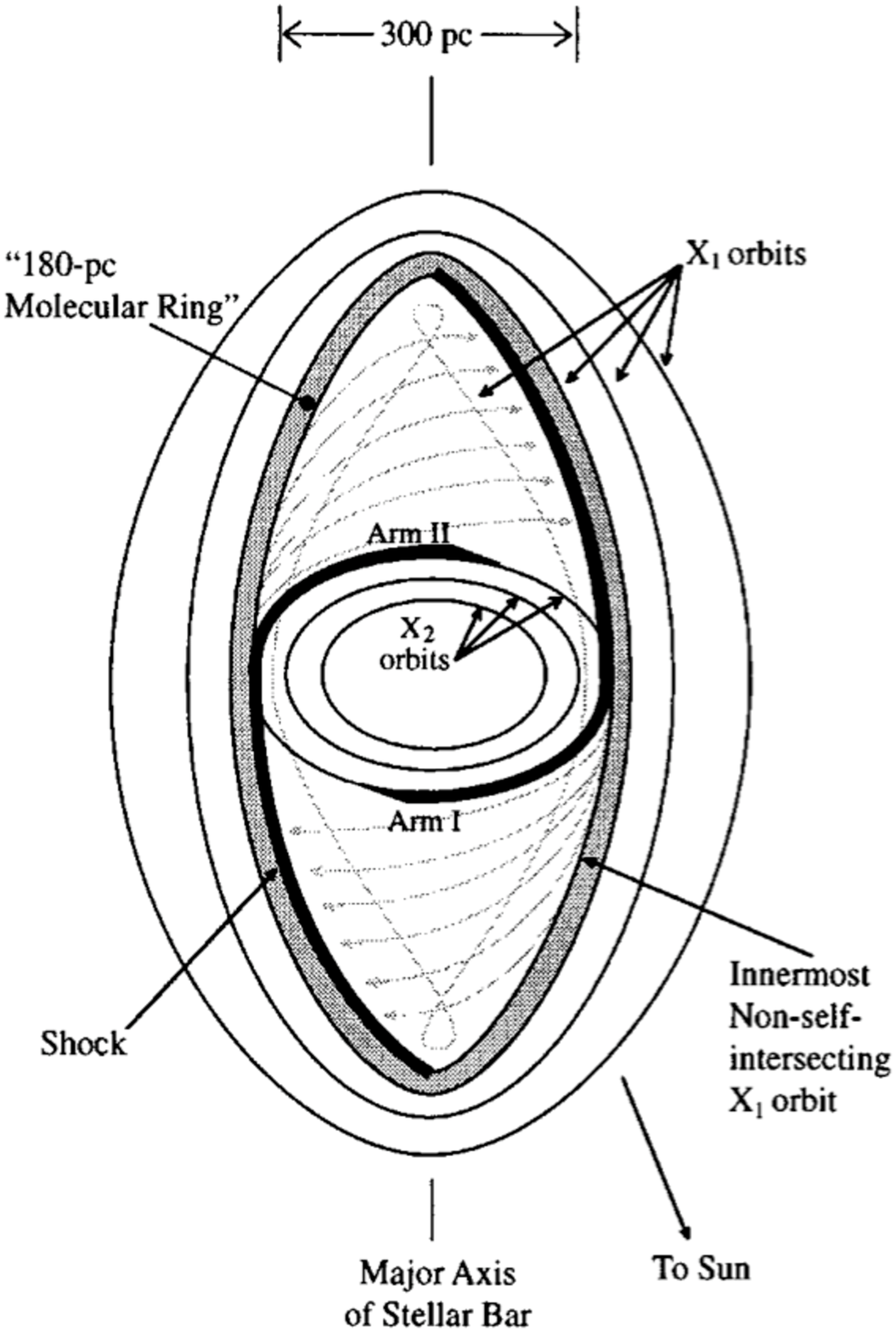}
\vspace{-7.85cm}

\hspace{+4.3cm}
\includegraphics[width=0.67\textwidth,height=0.72\textwidth,angle=-90]{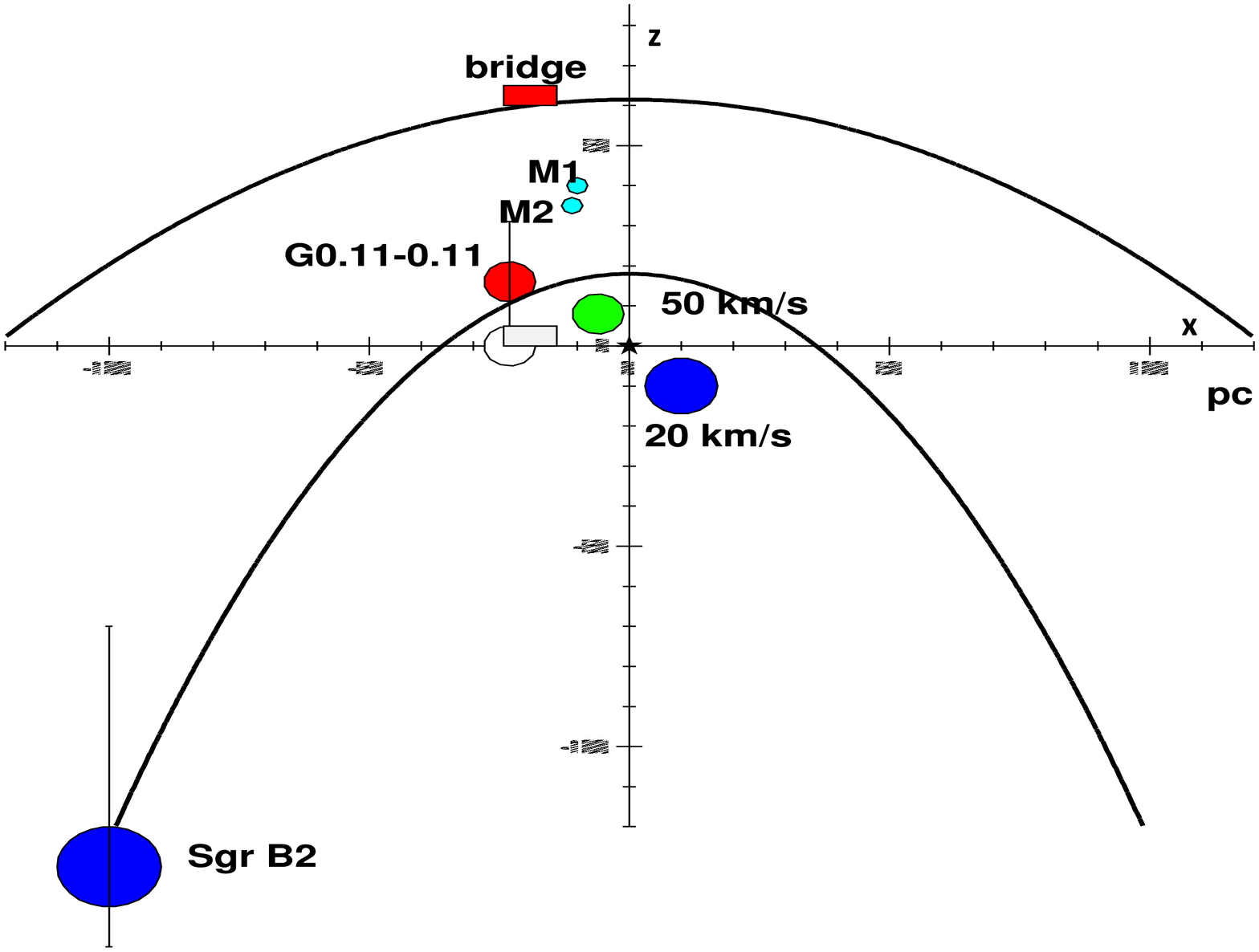}
\caption{{\it (Left panel)} Schematic diagram illustrating the shapes and relative 
orientations of the $X_1$ and $X_2$ orbits and the location of the shocks
resulting from the interaction between the innermost $X_1$ orbit and the 
outermost $X_2$ orbit. The arms in the CMZ hypothesized by Sofue (1995)
are also shown. [Figure courtesy of Morris \& Serabyn 1996].
{\it (Right panel)} Sketch of the Galactic plane as seen from the 
north Galactic pole. Sgr A* is at the vertex and Earth is toward the bottom. 
Circles represent the hypothesised locations of some MC of the CMZ. 
Parabolas represent hypothetical light fronts (as seen from Earth) emitted 
by Sgr A* 100 and 400 yr ago.
}
\label{CMZSketch} 
\label{CMZscan}
\end{figure}
The Central Molecular Zone (CMZ, see Fig. \ref{CMZ}) is a thin layer, 
of about $300\times50$ pc in size, that contains 
a total of about $\sim3-5\times10^7$ M$_{\odot}$ of dense ($n\sim10^4$ 
cm$^{-3}$), high filling factor ($f\gsimeq 0.1$) molecular material (Armstrong \& 
Barrett 1985; Walmsley et al. 1986; Bally et al. 1987; 1988; Gusten 1989; 
Stark et al. 1989; Tsuboi et al. 1989; Dahmen et al. 1998), which represents 
$\sim10$ \% of the 
total neutral mass content in the Galaxy (Gusten 1989; Morris \& Serabyn 1996). 
The MC in the CMZ differ considerably from those in the rest of the Galaxy, 
with higher gas temperatures ($T\sim70$ K; Gusten et al. 1981; Morris et al. 
1983; Mauersberger et al. 1986), highly supersonic internal velocities 
($\gsimeq15-50$ km s$^{-1}$) and higher densities ($n\sim10^4$ cm$^{-3}$). 
These characteristics are probably deeply related to the special location in 
the steep gravitational potential of the central bulge, to the GC environment 
and the past energetic GC activity. For example the high MC densities are probably
related to the large tidal shearing present in the CMZ. In fact, as a necessary 
condition for a cloud at a Galactocentric radius $R_G$ to be stable against 
tidal disruption, its mean density has to be: $n>10^4 cm^{-3} \times 
[75 pc/R_G]^{1.8}$ (Gusten \& Downes 1980). Any lower density MC 
would be sheared into a tenuous diffuse gas. 

The CMZ molecular gas is distributed in giant molecular cloud complexes, 
with three quarters of the dense clouds located at positive longitudes, 
three quarters have positive velocities and some have large radial and 
vertical motions (Bally 1988).
This distribution and kinematics are clearly inconsistent with both axial 
symmetry and uniform circular rotation (Bania 1977; Liszt \& Burton 1978; 
Morris et al. 1983; Bally et al. 1988; Jackson et al. 1996) and they 
suggest two different kinematic MC populations. 

\textbf{\emph{The disc population: }}
The most massive, lower velocity ($v\lsimeq100$ km s$^{-1}$) component 
(the so called "disc population", Bally et al. 1988; Heiligman 1987) resides 
close to the Galactic plane in the central 100-200 pc and has filament-like 
clouds with coherent velocity gradients, suggestive of dust lanes and tidally 
stretched arcs or arms of gas (Stark \& Bania 1986; Serabyn \& Gusten 1987; 
Bally et al. 1988; Sofue 1995).
We point out here that the CMZ disc population appears to have a toroidal 
geometry (see Fig. \ref{CMZ}; Molinari et al. 2011) and a physical scale 
remarkably similar to the 
outer parts of the AGN molecular torus invoked by AGN unification schemes 
(Antonucci 1993; Urry \& Padovani 1995). 

\textbf{\emph{180 pc ring: }} 
The second, higher velocity ($v\gsimeq130-200$ km s$^{-1}$) molecular gas 
component seems to be located in the outer boundary of the CMZ and it appears 
to form a quasi-continuous ring structure (the so called "180 pc ring") inclined 
by $\sim20^{\circ}$ with respect to the Galactic plane (Liszt \& Burton 1978; 1980). 
Further out the 180 pc ring connects with a larger structure with similar tilt, 
the so called "H I nuclear disc" that contains much of the neutral material 
($8\times10^6$ M$_{\odot}$; Bania 1977; Sofue 1995) lying between 
0.3 and 2 kpc from the nucleus (Liszt \& Burton 1978; Gusten 1989; Morris \& 
Serabyn 1996). Thus the outer part of the 180 pc ring appears to mark a 
transition between H I and H$_2$ (Binney et al. 1991).

\subsection{"The Expanding Molecular Ring"} 
\label{EMR}

The H I nuclear ring kinematics are highly perturbed, with outward 
motions of the same order as the rotation velocities. For this reason 
it was originally interpreted as a radially Expanding Molecular Ring or EMR 
(Scoville 1972; Kaifu 1972). 
Bania (1977) proposed that the expansion of this molecular ring was the 
result of an explosive event at the GC, which occurred $\sim10^6$ yr ago, 
providing the momentum impulse to $\sim10^7$ M$_{\odot}$ of nearby gas.
There were two main weaknesses of this interpretation. 
First is the rather extreme energy release required, being of the order 
of $E\sim10^{55}$ erg (Sanders 1989; Saito 1990), connected with the lack 
of evidence for the occurrence of such an energetic event. 
Secondly, the dearth of any clear evidence for interaction between the supposed 
flood of radially moving molecular material and the non-EMR clouds in the 
GC region (although Uchida et al. 1994a,b and Sofue 1995 found some 
indications for shocks as well as gaps in the EMR corresponding to the most 
prominent clouds in the CMZ). 
We point out here that the recently discovered \fermi\ bubbles (Su et al. 2010)
would require, if inflated by accretion onto Sgr A*, a similar energy release 
$E\sim10^{55}$ erg occurring on the correct time scale (a few $10^6$ yr ago). 
Thus part of the kinematical properties of the Expanding Molecular Ring 
might be carrying the vestige of a period of intense Sgr A* activity. 

However, we need to consider that at the distance of few $10^2$ pc 
from the GC (on the scales of the CMZ and EMR) 
the gravitational potential is dominated by the stellar component. 
Nowadays, it is well known that the Galactic bulge has a substantial stellar bar
(de Vaucouleurs 1964; Peters 1975; Sellwood 1993; Dwek et al. 1995; 
Zhao et al. 1994; Blum 1995) extending to at least 2.4 kpc and with a total 
mass of $1-3\times10^{10}$ M$_{\odot}$. 
Thus the kinematics of the EMR and CMZ are expected to deviate from 
axially symmetric circular motion (Binney 1994). In fact, gas moving in response to 
a bar potential tends to settle into elongated orbits (see the left panel of 
Fig. \ref{CMZSketch}). 
When the gas is orbiting at radii between that of corotation and that of the inner 
Lindblad resonance of the bar pattern, it moves on the so-called 
X$_1$ orbits (see Fig. \ref{CMZSketch},
Contopoulos \& Martzanides 1977), the long axis of which is aligned with the bar. 
There is an innermost stable X$_1$ gas orbit inside of which these orbits become 
self-intersecting, thus likely orbit-crossing and shocks will happen, leading 
to angular momentum loss and the gas settling into a new family of gas closed 
orbits, the X$_2$ orbits, oval-shaped orbits with long axes perpendicular to the bar. 
In this scenario the H I ring would correspond 
to the X$_1$ orbits, the 180 pc ring being the innermost stable X$_1$ orbit and 
the inner CMZ disc population the gas residing on the X$_2$ orbits (Binney et al. 
1991).
A 100 pc elliptical and twisted ring of molecular gas, consistent with residing 
in X$_2$ orbits, has been recently revealed by new $Herschel$ observations 
(Molinari et al. 2011), strengthening this interpretation (see Fig. \ref{CMZ}). 

Probably as a result of the tidal shearing and the associated cloud stability problem
(Gusten \& Downes 1980), the CMZ is composed of very clumpy cloud morphology, 
with about $\sim10$ \% of the mass in higher density clumps 
($\sim10^5$ cm$^{-3}$) embedded in a lower density inter-clump medium 
($\sim 10^{3.5}$ cm$^{-3}$).

\subsection{The CircumNuclear Disc, CND} 
\label{CND}

The inner 5 pc from Sgr A* contains a concentration (with total mass of 
$\sim10^{5-6}$ M$_{\odot}$) of dense and warm molecular and atomic 
gas (the so called CircumNuclear Disc, CND, 
Becklin et al. 1982; Genzel et al. 1985; Gusten et al. 1987; Christopher et 
al. 2005; Montero-Castano et al. 2009; Oka et al. 2011; 
Martin et al. 2012; Requena-Torres et al. 2012) with a configuration similar 
to a rotating disc and/or a set of filaments with a tilt of $20-30^{\circ}$ 
compared to the Galactic plane and very dense ($n_H\sim10^{6-8}$ 
cm$^{-3}$) clumps. The asymmetric CND distribution beyond its inner 
rim and the presence of also a component of high-velocity ($v\sim\pm300$ 
km s$^{-1}$) ionised gas, suggest either a transient origin of the CND 
(produced by e.g. a gravitationally captured, tidally stretched cloud; 
Quinn et al. 1985; Zhao et al. 1995), or an energetic disruption of a 
stable disc (e.g. produced by the impact of the Sgr A East supernova 
remnant upon the disc; Maeda et al. 2002; Rockefeller et al. 2005). 
The mass inflow rate of CND gas migrating into the central 
parsec is $\sim10^{-2}-10^{-4}$ M$_{\odot}$ yr$^{-1}$ (Gusten et al. 
1987; Jackson et al. 1993; Vollmer et al. 2002; Genzel et al. 2010), 
while the accretion rate onto the accretion shock (far less gas may 
actually make it to Sgr A* event horizon) is estimated to be few 
$\sim10^{-6}$ M$_{\odot}$ yr$^{-1}$ (Cuadra et al. 2008; Baganoff et al. 
2003; Xu et al. 2006). This strongly suggests that the present dim period 
in Sgr A*'s life might be just a temporary phenomenon.
The CND provides a reservoir of gas that can be accreted if local processes 
that cause angular momentum loss can bring material inwards. The ionised 
"arms" of Sgr A West (the so called mini-spiral) are possibly infalling fragments 
of the CND that can, on a time scale of $\sim10^{2-3}$ years, give rise to 
energetic accretion events. 

\subsection{The central cluster of young stars}
\label{youngstars}

Within the central parsec, there are several populations of stars. 
The nuclear cluster, 
forming 98 \% of the observed stars and composed of old ($>10^9$ yr), 
late-type giants and helium burning stars, plus $\sim200$ young 
($t=6\pm2\times10^6$ yr) stars (one of the richest massive star forming 
regions in the entire Galaxy), forming a strongly warped disc 
(Genzel et al. 2003; Paumard et al. 2006; Tanner et al. 2006; 
Lu et al. 2009; Gillessen et al. 2009; see Genzel et al. 2010 for an overview). 
The tidal shear of the Galactic black hole would impede normal star 
formation, so for a long time, the presence of the young stars around it gave 
rise to the "paradox of youth" (Ghez et al. 2003). 

It is now believed that they could have formed in situ (see Genzel et al. 2010 
for a review) in a disc either rapidly produced by a massive cloud plunging into 
Sgr A* (although it seems strange that a molecular cloud reaches the GC at 
small angular momentum) or slowly building up, but with star formation 
quenched, until a trigger is reached. 
Once the disc is formed, rapid dissipation and cooling of fragmented 
parts of the disc might occur. These clumps would then efficiently form 
stars (Nayakshin et al. 2007; Hobbs \& Nayakshin 2009; Gualandris et al. 2012). 
Monte Carlo simulations show that, in this case, the amount of gas accreted 
through the inner radius of the disc is large enough to provide 
Sgr A* with the gas to radiate near its Eddington limit throughout the star-formation 
episode. The ages of the young stars suggest that this event 
occurred $\sim6\pm2\times10^6$ yr ago, a time-scale compatible with the 
formation of the \fermi\ bubbles and the launch of the EMR. 
Simulations also show that a cloud captured by Sgr A* is 
expected to form a sheared and asymmetric circumnuclear disc very 
closely resembling the observed CND (although the likelihood of such an event 
needs to be demonstrated). Whichever the process that created it, the presence 
of a massive disc and star formation in the immediate vicinity of the black hole
would necessarily be accompanied by a high accretion rate onto Sgr A*. 
Therefore this star formation would be accompanied by strongly enhanced 
AGN activity (which, in fact, could have been the trigger for the 
star formation). Thus, the detailed study of Sgr A*'s environment reveals 
further hints of a possible glorious past for Sgr A*. 

\subsection{Fermi bubbles and evidence for Sgr A*'s jets}

A pair of large scale structures, extending up to 50 degrees (i.e. $\sim10$ kpc) 
above and below the Galactic plane and with a width of about $40^{\circ}$,
have been recently detected, at $\gamma$-ray energies, above a few GeV, 
thanks to \fermi\ data (Su et al. 2010). These so-called "\fermi\ bubbles" 
are centered on the core of the Milky Way, they have approximately uniform 
gamma-ray surface brightness with sharp edges. The bipolar morphology 
and sharp edges suggest that the \fermi\ bubbles originated by some large 
episode of energy injection in the GC (e.g., jets originating from AGN activity
or nuclear starburst). The bubbles have a luminosity of 
$4\times10^{37}$ erg s$^{-1}$, they extend to $\sim10$ kpc outside the 
Galactic disc and have an estimated energy content of the order of 
$\sim10^{55}$ erg (Bland-Hawthorn \& Coen 2003; Su et al. 2010). 
The \fermi\ bubbles have a hard spectrum extending up to 100 GeV and they 
seem to be associated with large scale structures observed in X-rays with \rosat\ 
(Bland-Hawthorn \& Cohen 2003) and with an excess of radio emission 
found in \wmap\ data (the so called Galactic haze; Dobler \& Finkbeiner 2008). 
New \planck\ data have recently confirmed the microwave haze to have a 
morphology consistent with the \fermi\ bubbles. The \planck\ data reveal 
a hard microwave spectrum consistent with hard synchrotron radiation and 
excluding free-free emission (Planck collaboration 2012). 

The mechanism producing the bubbles is still quite uncertain. 
Su et al. (2010) discussed a possible starburst phase in the GC a few million 
years ago. A modest, but constant, injection of $\sim10^{39}$ erg s$^{-1}$ of 
cosmic rays to the halo is today inferred by the IR luminosity and $\gamma$-ray 
emission from the CMZ (Crocker et al. 2011). Integrating this emission over 
$\sim10^9$ yr would produce the correct energy budget (Crocker \& Aharonian 
2011), however a mechanism to confine the particles for such a long time would 
be required. Alternatively, the \fermi\ bubbles, X-ray structure and microwave haze 
might all be produced by an intense AGN phase, with Sgr A* accreting close to the 
Eddington limit, a few $10^6$ yr ago (Zubovas et al. 2011). 
Recently Su \& Finkbeiner (2012) reported a gamma ray cocoon feature, in the 
southern \fermi\ bubble, and they produced evidence for a jet-like feature along 
the cocoon axis with a luminosity of 
$L_{\rm 1-100~Gev}\sim5.5\pm0.45$ and $1.8\pm0.35 \times 10^{35}$ 
erg s$^{-1}$ for the cocoon and jet, respectively. If this finding is confirmed, 
this would strongly support 
an accretion phase onto Sgr A* that would be therefore associated with a 
relativistic jet and outflows/winds as generally observed in AGN and 
stellar mass black holes (Antonucci 1993; 
Fender et al. 2004; Ponti et al. 2012). For deeper insight into the physics 
associated with the \fermi\ bubbles we point the reader to the contribution, 
in this same volume, by M. Su and R. Crocker. 

On a much smaller scale of a few parsec, a faint, collimated linear structure 
observed at radio wavelength has been recently claimed by Yusef-Zadeh et 
al. (2012). This feature seems to arise from Sgr A* and to interact with the 
ionised and molecular material orbiting the supermassive BH. 
Yusef-Zadeh et al. (2012) associate this linear structure to a jet possibly 
originating a few hundred years ago. Deeper observations are necessary to 
confirm this feature. 
A different linear feature pointing to the location of Sgr A* has been 
noted in X-ray images of the central parsec (Muno et al. 2008; Li et al. 2012). 
The X-ray emission is interpreted as being synchrotron emission produced 
by a population of highly relativistic electrons created by the shock 
at the collision point between the jet and the Eastern Arm of the mini-spiral
inside the CND (Li et al. 2012).

\section{Is the GC Fe K$\alpha$ emission the echo of a past Sgr A* flare?}

\begin{figure}[t]
\hspace{-1.3cm}
\includegraphics[scale=0.35,angle=0]{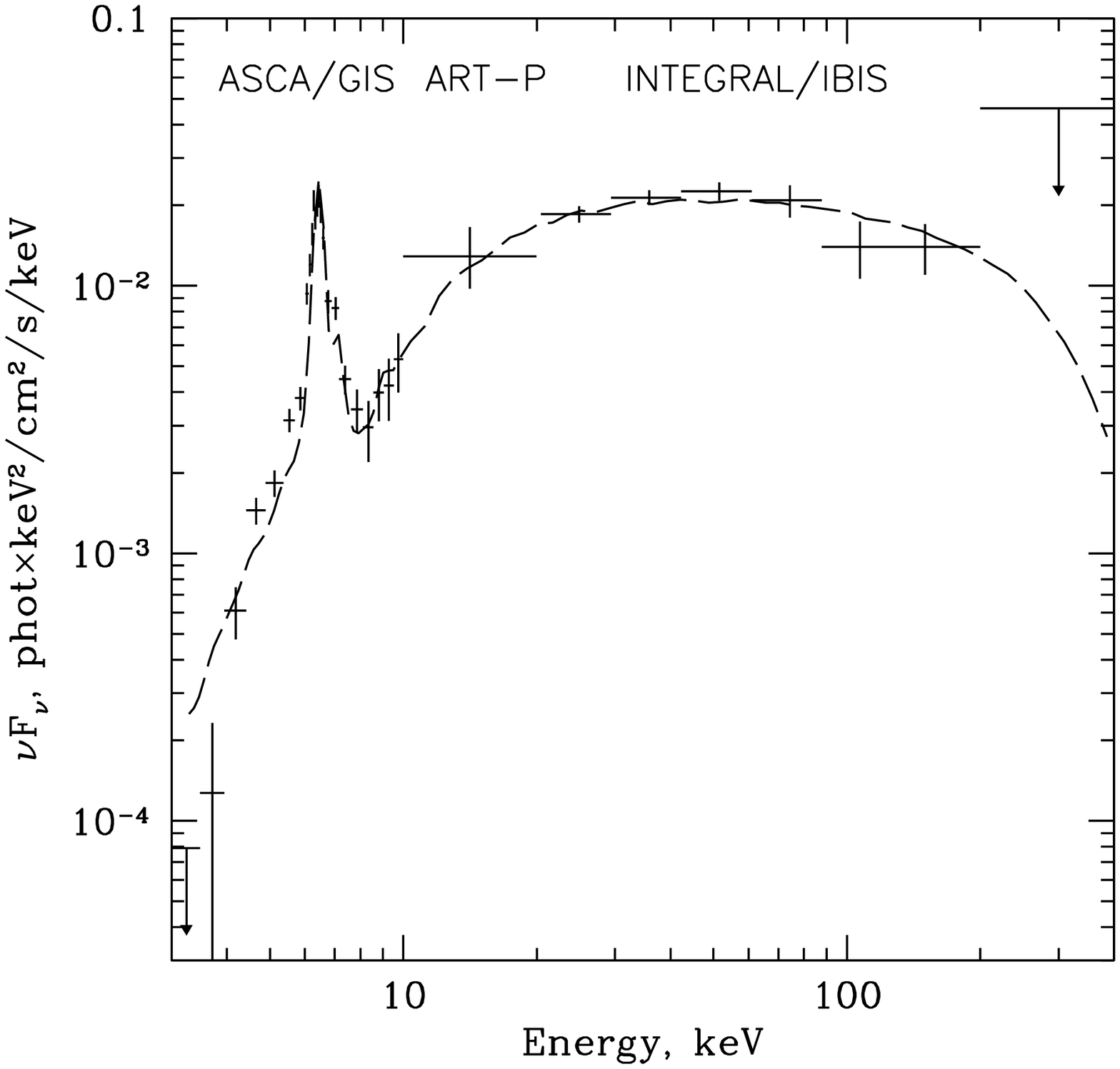}

\vspace{-7.9cm}

\hspace{+5.5cm}
\includegraphics[width=0.65\textwidth,height=0.63\textwidth,angle=0]{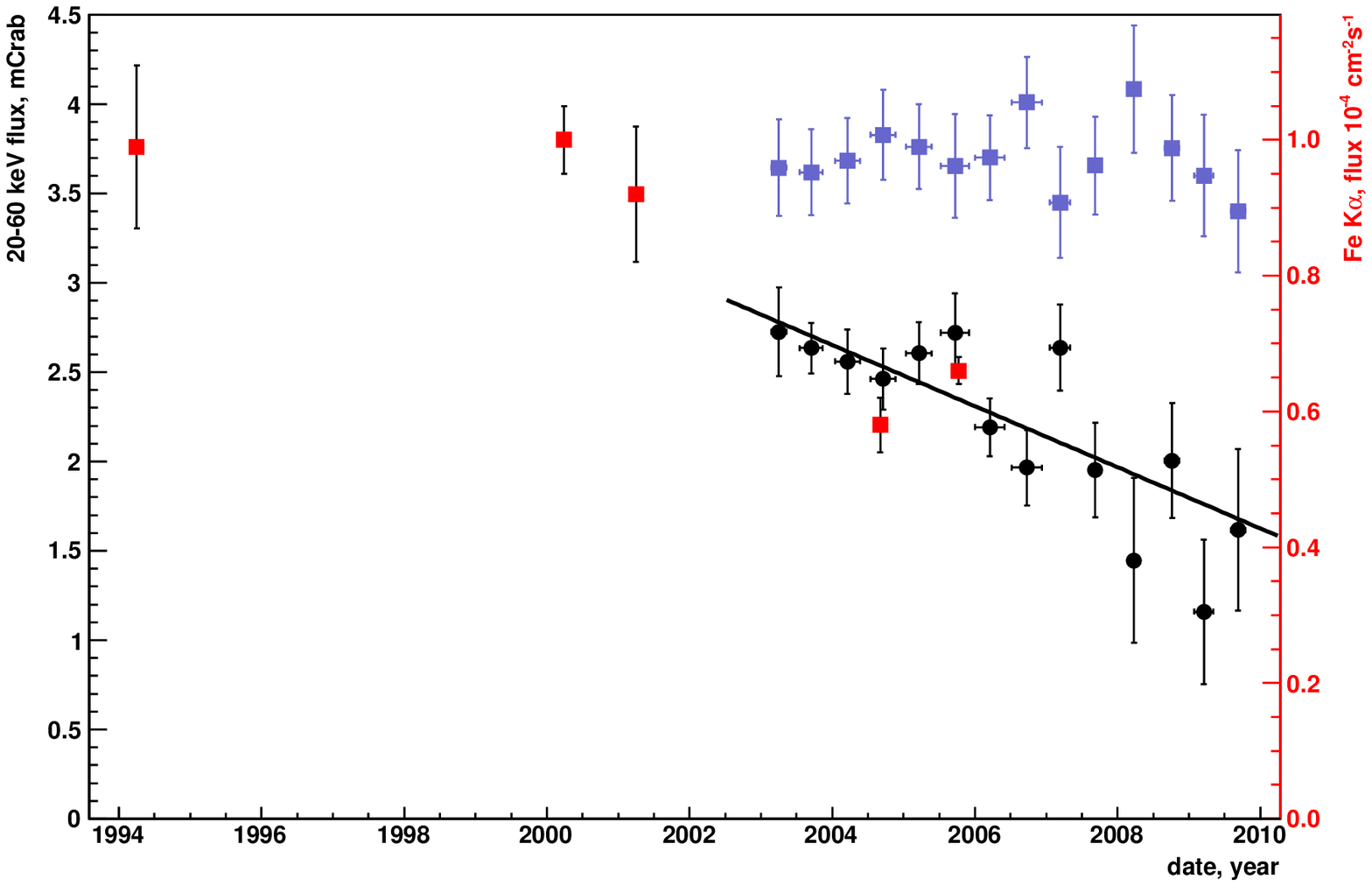}
\caption{{\it (Left panel)} \asca, $ART-P$ and \integral\ broad band 
X-ray spectrum of Sgr B2 (1 $\sigma$ error bars and 2 $\sigma$ upper
limits). The dashed line shows the best fit reflection model convolved with 
the instrumental resolution. [Figure courtesy of Revnivtsev et al. (2004)]
{\it (Right panel)} \integral\ light curve of the corrected 20-60 keV Sgr B2 
flux in mCrab (black circles) and of the secondary calibrator, Ophiuchus cluster 
(light blue squares). Superimposed (red squares) are the Fe K$\alpha$ line 
fluxes from Inui et al. (2009). A chi-squared test favors a linearly decreasing flux 
over a constant at the $\sim4.8 \sigma$ level. [Figure courtesy of Terrier 
et al. (2010)]
}
\label{SgrB2} 
\end{figure}
The discovery of the Fe K$\alpha$ line and hard X-ray emission from several 
massive MC in the CMZ placed on solid ground the idea of this 
emission is being produced by reflection of a past GC flare (Sunyaev et al. 1993; 
Koyama et al. 1996). Deep \asca\ observations of Sgr B2, the most massive 
MC complex in the CMZ, showed many spectral features reminiscent of an 
X-ray reflection nebula (see Fig. \ref{SgrB2} and \ref{Refl}), such as: 
i) strong and extended Fe K$\alpha$ emission with $\sim1-2$ keV EW; 
ii) a low-energy cut-off below $\sim4$ keV; iii) a sharp 
flux drop consistent with the presence of an Fe K edge (Koyama et 
al. 1996; Murakami et al. 2000). Not only spectral, but also morphological
evidences has been accumulating. For example, in the Sgr B complex, 
Murakami et al. (2000) observed 
the peak of the extended 6.4 keV emission from Sgr B2 being shifted toward 
the GC by about $\sim1-2$ arcmin compared to the molecular mass 
distribution, thus suggesting the presence of an irradiating X-ray source 
located in the direction of the GC. Assuming that Sgr A* is the illuminating source, 
they estimated that, a few hundred years ago, it had to be orders of magnitudes 
brighter than now, with a 2-10 keV luminosity, 
$L_{\rm 2-10}\sim3\times10^{39}(d/100 pc)^2$ erg s$^{-1}$.  
A subsequent long \chandra\ observation confirmed these findings (allowing also 
the detection of the weaker Fe K$\beta$ emission line) and finally, owing to the 
excellent imaging resolution, the \chandra\ data ruled out the possibility that the 
6.4 keV emission was due to unresolved point sources, which contribute 
less than $3$ \% of the diffuse Fe K$\alpha$ emission (Murakami et al. 2001). 

The combination of \integral, \asca\ and \granat\ data on Sgr B2, allowed the 
detection, for the first time, of the Compton hump and the characterisation of 
the broad band 2-200 keV spectrum (see Fig. \ref{SgrB2}), which is well 
fitted by scattered and reprocessed 
radiation in a cloud of cold gas with abundances about twice Solar (Revnivtsev et al. 
2004). Moreover, Revnivtsev et al. (2004) found no significant variability of the 
Fe K$\alpha$ during the 1993-2001 period. They excluded an internal 
source hypothesis and, based on the intensity of the broad-band spectrum and 
the cloud column density, they estimated a high luminosity, 
$L_{\rm 2-10 / 2-200}\sim0.5/1.5\times10^{39} (r/10 pc)^{-2} (d/100 pc)^2$ 
erg s$^{-1}$ of the primary illuminating source, active for many years. 
Such luminosity and period of activity are higher (even higher than 
the Eddington luminosity for M$_{\rm BH}=10$ M$_{\odot}$)
and much longer than the typical values observed in X-ray binaries, 
thus disfavouring these as possible candidates.

Fe K$\alpha$ emission has also been detected in a number of other MC, 
such as Sgr C (Murakami et al. 2001; Sidoli et al. 2001) and small regions 
in the Sgr A complexes, such as the G0.11-0.11 cloud located close to the radio arc 
and the bridge (Bamba et al. 2002; Park et al. 2004). All these Fe K$\alpha$ clumps 
are associated with molecular complexes and they have 
EW$_{\rm Fe K\alpha}\sim1-2$ keV, in agreement with a reflection origin. 
The increased \suzaku\ sensitivity allowed Koyama et al. (2007a) to detect 
also the Ni K$\alpha$ emission from several molecular complexes and 
to discover, in the Sgr B complex, two other X-ray nebulae, M0.74-0.09 
and M0.51-0.10 (the latter coincident with Sgr B1). These molecular complexes 
show a decay of the Fe K$\alpha$ emission similar to the one observed in 
Sgr B2, moreover if illuminated by Sgr A* they would require similar luminosity,
suggesting that they might be illuminated by the same source irradiating Sgr B2 
(Koyama et al. 2007b; Nobukawa et al. 2008; but see also Yusef-Zadeh et al. 2007). 

\subsection{Alternative mechanisms to produce FeK$\alpha$ emission from MC}

X-ray irradiation can induce Fe K$\alpha$ line emission by removal of a 
K-shell electron rapidly followed by an electronic transition from the L shell 
to fill the vacancy (see \S \ref{refl}). However, the same process 
can also be produced by collisional ionisation induced by accelerated particles, 
such as low-energy cosmic ray electrons as well as protons/ions. 
Moreover, the energetic particles, diffusing in dense neutral matter, do produce 
nonthermal X-rays by atomic collisions as they slow down by ionisation 
and radiative energy losses (non-thermal bremsstrahlung), thus producing 
a hard X-ray continuum emission (Valinia et al. 2000; Tatischeff 2003; 2012). 

The physical conditions in the GC are exceptional (see \S \ref{GCenvironment})
and make it a special place as regard to particle acceleration. 
Bright non-thermal radio filaments trace magnetic fields up to a mG (Ferriere 2009)
and large-scale non-thermal emission suggest the average field to be at least 50 
$\mu$G (Crocker et al. 2010). Three of the most massive young star clusters
in the Galaxy lie in the CMZ, and intense star formation is taking place there 
at several sites. It is possible that the density of low-energy cosmic rays is 
larger in the central regions than in the rest of the Galaxy (Berezinskii et al. 
1990; Dogiel et al. 2002). 

{\bf Cosmic ray electrons:} 
The presence of strong non-thermal radio filaments in the vicinity of several MC 
(Yusef-Zadeh et al. 2002; 2007) makes it reasonable to consider the presence 
of a large density of cosmic ray electrons there. This is also supported by the 
observation of diffuse low-frequency radio emission in the GC region (LaRosa 
et al. 2005) and the high estimates of the ionisation rate compared to the values 
obtained in the Galactic disc (Oka et al. 2005). It has also been suggested 
(Yusef-Zadeh et al. 2007) that the heating of MC induced by the 
interaction with cosmic ray electrons might explain the long-standing problem 
of the origin of the high temperature of the MC within the CMZ (Morris \& 
Serabyn 1996).
Several authors (Valinia et al. 2000; Tatischeff 2003; 2012) studied in detail 
the production of non-thermal lines and X-ray continuum by the interaction of 
cosmic ray electrons with neutral ambient gas. In most cases the EW of the 
Fe K$\alpha$ line is found to be lower than $\sim0.3-0.5\times
(A_{\rm Fe}/A_{\odot})$ keV (where $A_{\rm Fe}/A_{\odot}$ is the Fe fractional 
abundance compared to Solar) and never higher than $\sim0.6\times
(A_{\rm Fe}/A_{\odot})$ keV, and the continuum radiation should be generally 
hard ($\Gamma<1.4$). The production of Fe K$\alpha$ photons is relatively 
inefficient. The Fe K$\alpha$ total luminosity from the CMZ is 
$\gsimeq6\times10^{34}$ erg s$^{-1}$ (Yusef-Zadeh et al. 2007), thus 
requiring a power $\gsimeq2\times10^{40}$ erg s$^{-1}$ in cosmic ray electrons 
(assuming Solar metallicities), which is comparable to that contained in cosmic 
ray protons in the entire Milky Way (Tatischeff et al. 2012). 

{\bf Cosmic ray ions/protons:}
Alternatively, subrelativistic ions propagating through MC can radiate (through 
inverse bremsstrahlung from the fast ions and classical bremsstrahlung 
from the secondary knock-on electrons) in the hard X-ray domain and create 
Fe K$\alpha$ vacancies (Dogiel et al. 2009; 2011; Tatischeff et al. 2003; 2012). 
Dogiel et al. (2009) proposed that the tidal disruption and subsequent accretion 
of stars by the central BH (suggested to happen at a rate of one every $\sim10^4$ yr; 
Syer \& Ulmer 1999) could produce the required rate (basically constant 
on time-scales of centuries) of subrelativistic protons to induce the observed 
Fe K$\alpha$ emission. 
For strong shock acceleration of non-relativistic particles and strong particle 
diffusion in the cloud, the expected Fe K$\alpha$ EW is expected to be 
$\sim0.6-0.8\times(A_{\rm Fe}/A_{\odot})$ keV, with an associated continuum 
having spectral photon index $\Gamma\sim1.3-2$. However, fast ions with 
a soft source spectrum can produce very large EW ($\gg1$ keV) and 
steep ($\Gamma\gg2$) power-law continuum (Dogiel et al. 2011; Tatischeff 
et al. 2012). In this scenario the Fe K$\alpha$ line width is expected 
to be several tens of eV, about one order of magnitude wider than 
for X-ray reflection or cosmic ray electron scenario 
(Tatischeff et al. 1998; Dogiel et al. 2009). In fact, the X-ray lines produced 
by collisions of ions heavier than 4He can be shifted by several tens of eV,  significantly broadened and split up into several components (Garcia et al. 
1973; Tatischeff et al. 1998). For example, the Fe K$\alpha$ line produced by 
impacts of O ions of 1.9 MeV/nucleon is blueshifted by $\sim50$ eV in 
comparison with the one produced by 5-MeV proton impacts, and has a 
FWHM of $\sim100$ eV (see Garcia et al. 1973, Fig. 3.55)

As another alternative, Bykov (2003) proposed that subrelativistic particles are 
produced by fast-moving knots resulting from supernova explosions. 
In fact, fast moving supernova ejecta interacting with dense MC provide 
a fast conversion of kinetic energy into IR and X-ray radiation with a hard 
($\Gamma\sim0-1.5$) spectrum and Fe K lines (6.4 and 6.7 keV) with 
total EW$\sim0.5-0.6$ keV. Just $\sim3$ supernova remnants of age 
less than $\sim10^3$ yr in the GC region could provide the required 
number of fast moving knots (Bykov 2003).

\subsection{A Christmas tree around Sgr A*}

A major step forward has been the detection of variability from Fe K$\alpha$
emitting clouds. The first such detection was reported by Muno et al. (2007), 
who showed a $\sim2-4~\sigma$ continuum variability from filamentary 
($\sim0.3\times2$ pc) regions that are dominated by Fe K$\alpha$ emission and 
coincident with molecular structures near Sgr A (Muno et al. 
2007). Fast variability is a key factor to discriminate X-ray reflection nebulae 
from Fe K$\alpha$ emission induced by cosmic ray interactions with MC 
(Valinia et al. 2000; Yusef-Zadeh et al. 2002; 2007; Bykov 2002; Dogiel et al. 2009). 
Fast Fe K$\alpha$ or hard X-ray continuum variability, in fact, rules out the 
cosmic ray proton scenarios and is barely consistent with the cosmic ray 
electron models.
The variations observed by Muno et al. (2007) in the Sgr A complex could be 
produced by an X-ray binary of 
$L_X\sim10^{37}$ erg s$^{-1}$ at a distance of $\sim7$ pc. 
If, instead, the irradiating source is Sgr A* (thus at, or further away than, the projected 
distance of 14 pc) then $L_X\gsimeq10^{38}$ erg s$^{-1}$. Muno et al. (2007)
suggested that the flare illuminating these MC in Sgr A might be different 
from the one irradiating Sgr B2. 
If these MC were located at their projected distances, such an outburst, 
in fact, would have occurred $t\gsimeq60$ yr or $t\sim20$ yr ago, before the 
advent of wide field X-ray monitors, thus it could have been missed (Muno et al. 
2007). 

Collecting data from \asca, \suzaku, \chandra\ and \xmm, Inui et al. (2009) and 
Koyama et al. (2008) observed a decline, of the order of $\sim60$ \% in $\sim10$ yr, 
of the Fe K$\alpha$ line flux from both the Sgr B2 and Sgr B1 regions. 
Using \integral\ data, Terrier et al. (2010) measured a $4.8 \sigma$ variation 
of the 20-60 keV (the energy of the Compton hump) flux from Sgr B2, starting 
to decline after the year 2000 (black points in Fig. \ref{SgrB2}). 
This confirmed, without inter-calibration effects, the Fe K$\alpha$ decay (red 
points in Fig. \ref{SgrB2}) previously observed. 
The observed decay time ($t_d=8\pm1.7$ yr) is 
compatible with the light crossing time of the MC core, consistent with reflection 
but inconsistent with an electron cosmic ray scenario. 
The spectral index of the illuminating 
power-law source is found to be $\Gamma=2\pm0.2$. Parallax measurements 
of the distance of Sgr B2 place it (assuming it is on a nearly circular Galactic orbit)
$130\pm60$ pc closer to us than Sgr A* 
(Reid et al. 2009; Ryu et al. 2009; but see also Molinari et al. 2011). 
This suggests a mean luminosity for the Sgr A* outburst of 
$L_X=1.5-5\times10^{39}$ erg s$^{-1}$, occurring 
$t=100^{+55}_{-25}$ yr ago (Terrier et al. 2010). 
New \suzaku\ observations obtained in fall 2009, when compared to previous 
observations obtained in fall 2005, confirm Sgr B's fading (Terrier 
et al. 2010) detecting, in 4 yr, a reduction in Fe K$\alpha$ flux and hard (8-10 keV) 
continuum by a factor 1.9-2.5 from both Sgr B2 and M0.74-0.09 
(Nobukawa et al. 2011). In particular the variations of these two molecular 
cores decrease in a synchronised way. The correlation of the variability between 
the two MC indicates that they share a common external reflection origin 
(Nobukawa et al. 2011). 

The Sgr A and Sgr B complexes are not the only ones to display 
Fe K$\alpha$ variations. 
The comparison between a \suzaku\ observation in January 2006 and  
previous \asca\ and \chandra\ observations suggested variability of the 
different Fe K$\alpha$ clumps in the Sgr C complex, all having 
EW$_{\rm Fe K\alpha}\sim2.0-2.2$ keV and clear Fe K$\beta$ (Nakajima et al. 2009). 
In particular, the clump M359.38-0.00 was not previously observed to emit 
Fe K$\alpha$ photons. Moreover, the Fe K$\alpha$ emission from M359.43-0.07 
appeared shifted compared to previous \asca\ observations, suggesting 
a reflection origin of this emission (Murakami et al. 2001). 
Interestingly, 
the M359.38-0.00 and M359.47-0.15 clumps correspond to molecular 
complexes with different line of sight velocities, which are thought to belong 
to two different arms of the CMZ MC population (see Fig. \ref{CMZSketch};
Sofue 1995; Sawada et al. 2004; Nakajima et al. 2009). This 
indicates that they most probably have very different line-of-sight distances 
compared to Sgr A*. 
This seems to suggest that these two clumps are irradiated by two different 
beams of radiation, for example by two different flares or by two phases of 
the same long flare. 

\begin{figure}[t]
\begin{center}
\includegraphics[scale=0.62,angle=0]{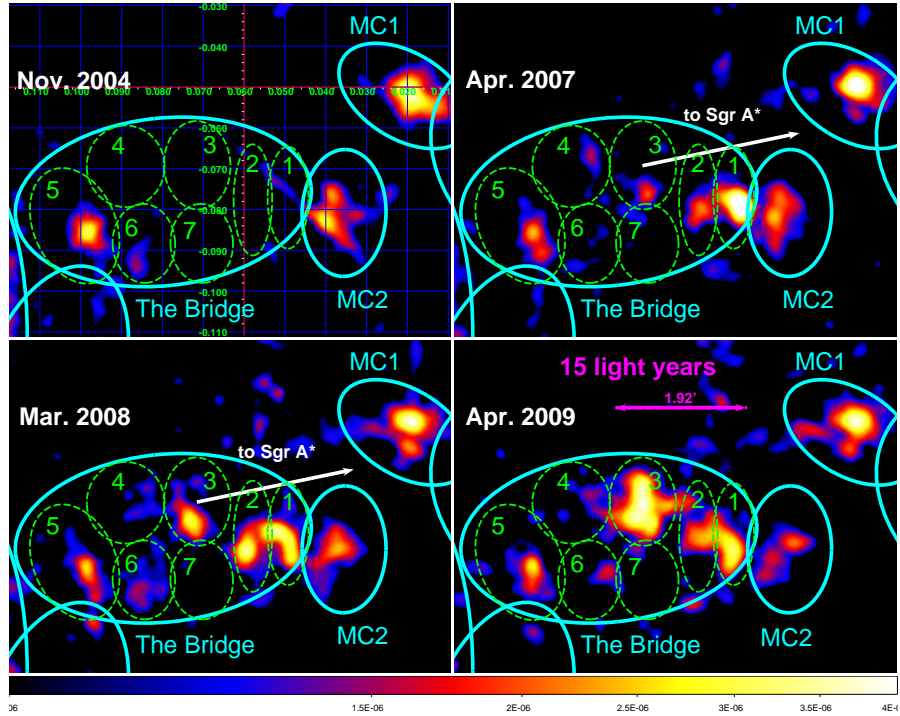}
\caption{Fe K$\alpha$ continuum-subtracted mosaic image of the 
different EPIC-pn observations of the bridge region. 
A brightening of regions 1, 2, 3 and 4 is clear. Such variation 
occurs in a time-scale of about 2-4 years, but on a spatial scale of about 
15 light years. This apparent superluminal motion can be explained if the 
bridge MC is illuminated by a bright ($L>1.3\times10^{38}$ erg s$^{-1}$) 
and distant ($>15$ pc) X-ray source active for several years. The observation 
of the superluminal echo cannot be explained by either a single internal source 
or by low energy cosmic ray irradiation. It is also highly unlikely that the variation
is produced by several uncorrelated sources. We note that the illumination starts 
in the Galactic west and propagates to the east, suggesting that the source is 
located in the direction of Sgr A*. }
\label{Bridge}
\end{center} 
\end{figure}
The long (more than 1 Ms exposure) \xmm\ monitoring campaign of the 
Sgr A complex (lasting for almost a decade), led to another 
important step forward. It allowed the first detection of highly significant 
($\sim4-13 \sigma$) 
Fe K$\alpha$ variations from several MC in Sgr A (Ponti et al. 2010) as well 
as the detection of a superluminal echo moving with apparent velocity 
of $v_{\rm app}\sim3$ c (see Fig. \ref{Bridge}) from a molecular complex 
called the bridge (Gusten Downes 1980). 
The superluminal propagation of the Fe K$\alpha$ emission strongly suggests 
an external illuminating source and rules out models based on internal 
sources and/or cosmic rays (which would have to deal with superluminal motions!).

Contrary to the case of the Sgr B complex, where every clump is consistent with 
being produced by the same flare, the Fe K$\alpha$ emitting regions in Sgr A 
have very different light curves (Ponti et al. 2010; Capelli et al. 2012). 
In some MC the Fe K$\alpha$ emission is rising, in others it is fading and in some 
it is constant (Ponti et al. 2010; Capelli et al. 2012). It is interesting to note that, 
as in Sgr B2, the Fe K$\alpha$ emission from G0.11-0.11 (a MC located 
close to the radio arc) is fading on a short time scale (comparable to the light crossing 
time of the MC). If located at its 
minimal distance from Sgr A* (25 pc) it would require $L_X\gsimeq10^{39}$ 
erg s$^{-1}$, $t\gsimeq75$ yr ago (Ponti et al. 2010). These estimates would 
necessitate a flare of Sgr A* with a luminosity and a difference in time of the same order 
of magnitude as the one implied by Sgr B2 and B1. Moreover 
all three of these MC are fading at a rate comparable to the MC light crossing time.
This seems to suggest that Sgr B1, B2 and G0.11-0.11 are illuminated 
by the same flare.
On the contrary the bridge (see Fig. \ref{Bridge}) shows a clear ($13 \sigma$) 
increase of Fe K$\alpha$ emission, while other MC are consistent with being 
constant over the \xmm\ monitoring (Ponti et al. 2010). 
While the flare seems to be passing out of some clouds, it is entering others, 
or is still in the act of passing through others. The rise, fall, and dwell time 
scales will all be important for reconstructing the flare profile, and for 
deciding whether there have been multiple flares within the past several 
hundred years.

\subsection{Which source of irradiation?}
\label{whichsource}

The very different variability behaviour might indicate that each one of these 
MC is responding to a different flare of, possibly, separate sources. 
The observation of a superluminal echo (Ponti et al. 2010) and the synchronised 
fading of the causally disconnected regions in the Sgr B complex (Nobukawa et al. 
2011) constrain the mechanism of Fe K$\alpha$ production to be irradiation 
from a powerful external source. 

As described in \S \ref{X-rayemission}, only 3 sources in the central 1.2 square 
degrees have reached luminosities higher than L$_X\gsimeq10^{37}$ erg 
s$^{-1}$ during the past few decades (Degenaard et al. 2012; Wijnands et al. 
2006). The highest peak luminosities have been reached by 1A 1742-289 
and GRO J1744-28 with L$_X\sim7\times10^{38}$ erg s$^{-1}$ 
and L$_X\sim3\times10^{38}$ erg s$^{-1}$, respectively (Degenaard et al. 2012). 
Both these sources are X-ray transients, thus active only for few weeks-months. 
No known X-ray binary undergoes such a long outburst with such 
high mean luminosity (Coriat et al. 2012) as the source irradiating 
the Sgr B complex. 
However, this does not exclude 
that bright X-ray transients, located close enough to a MC, might produce 
detectable Fe K$\alpha$ echos. Moreover, it does not exclude a past 
outburst of an even more extreme version (with even longer recurrence 
time and higher luminosity) of the peculiar BH X-ray binary GRS1915+105 
(a source accreting close to the Eddington rate for more than a decade; 
Fender \& Belloni 2004). Although this is possible, we believe that a flare 
from Sgr A* is more plausible and more parsimonious for explaining all of
the variability throughout the GC region. 

Alternative possibilities involve bright X-ray sources produced by either the 
passage of the shock wave, generated by the supernova explosion that gave 
birth to the Sgr A East remnant, over the supermassive BH (Maeda et al. 2002; 
Rockefeller et al. 2005) or the interaction between Sgr A East and the 
50 km s$^{-1}$ MC (Fryer et al. 2006). 
Fryer et al. (2006) performed three-dimensional hydrodynamic 
simulations of the evolution of Sgr A East and its interaction with surrounding medium 
(e.g. Sgr A* and the 50 km s$^{-1}$ MC). They found that the passage of the remnant 
across Sgr A* would have enhanced the accretion rate onto the BH by less than a 
factor of 2, thus not enough to explain the Fe K$\alpha$ emission from the 
CMZ. On the other hand, the first impact of the shock wave with the material 
in the 50 km s$^{-1}$ MC would have produced a luminosity of 
$L_{\rm 2-200~keV}\sim10^{39}$ erg s$^{-1}$, decreasing to $\sim10^{36}$ erg 
s$^{-1}$ in the following few centuries. 
The resulting light curve would have, therefore be predicted to have a sharp rise, 
followed by a very long decay time (a few centuries), dictated by the time-scale 
for dissipation of kinetic energy (Fryer et al. 2006). 
In this scenario the observed rapid and small amplitude Fe K$\alpha$ variability 
might be produced by the interaction of the Sgr A East shock with dense matter 
clumps in the 50 km s$^{-1}$ MC. 
However, it seems difficult to reconcile the expected slow luminosity 
decay with the fast decrease of a factor of $\sim3-4$ in 5-7 yr 
(Terrier et al. 2010; Nobukawa et al. 2011) observed in Sgr B2.

Before attempting to reconstruct the past activity in the GC, 
we summarise the main limitations and uncertainties associated 
with the tools that are used.

\subsubsection{Uncertainties on the luminosity estimates}

We try to estimate here the dominant source of uncertainty in the estimation 
of the primary source's past luminosity. Equation \ref{eq} (which assumes isotropic 
source emission and a cloud with $\sigma_T\ll1$) provides a direct link between 
the observed Fe K$\alpha$ line flux (or upper limit) and luminosity of the source
(e.g. Sgr A*), and is the main tool. 
The luminosity of the source depends linearly on the Fe K$\alpha$ line flux and 
inversely on the distance to the GC. Both are generally well known. 
The former to better than $10-15$ \% and the latter to less than 5 \%. 
A higher uncertainty is associated with the knowledge of the Fe abundance, 
$\delta_{Fe}$, which however can be constrained from the Fe K$\alpha$ 
line EW and is expected to have fluctuations of a factor of less than a few. 
Larger uncertainties are associated with the determination of the cloud 
optical depth, $\tau_T$, and mainly the solid angle $\Omega$ of the cloud 
from the location of the primary source. 

{\bf Solid angle:} This latter parameter ($\Omega=A_{\rm cloud}/4\pi d^2$) depends 
on the ratio between the area exposed to the illuminating source 
($A_{\rm cloud}$) and the square of the distance ($d$) between the source 
and the cloud. Fig. \ref{CMZ} shows a large concentration of molecular matter 
along any direction of the CMZ. In particular, several clouds typically overlap
along the line of sight, thus confounding the boundaries between the clouds. 
Each cloud is expected to move with approximately the same velocity. 
Thus, thanks to the molecular cloud data cubes that show the MC emission 
in each velocity range, we can separate the emission from each cloud along 
the line of sight and accurately measure the correct MC area.
MC typically have complex 
shapes (see Fig. \ref{CMZ}) and the area irradiated by the source might be 
different from the area projected on the image, however we expect 
this difference to be smaller than a factor of a few. 

{\it The dominant uncertainty is associated with the line-of-sight component
of the distance between the cloud and Sgr A*, to be $d_l$.} 
Being the CMZ compact, we expect that for each CMZ cloud, $d_l$ 
is constrained between the projected distance and $\sim\pm300$ pc. 
This generally implies a very large uncertainty that can be as high as 
1-2 orders of magnitude. 
Thus, knowledge of the line of sight distance between the MC and Sgr A*'s 
plane is key for accurate luminosity estimates. 
As discussed in \S \ref{GCenvironment} there are several indications of 
coherent patterns in the MC distributions, from which distances might be 
derived. However, although these large molecular 
structures incorporate a significant part of the gas, 
a further effort is required in order to accurately associate each 
molecular clump with a line of sight distance. 
Other methods to determine $d_l$ are based on extinction measurements. 
Sources distributed uniformly around the GC will appear more extinguished if the MC 
in question is placed more in the foreground, while MC placed at the far side 
of the CMZ will contribute little or no absorption. 

Ryu et al. (2009), assuming that the 6.7 keV and the diffuse soft X-ray  
plasma (see \S \ref{X-rayemission}) emission are uniformly distributed 
around the GC, measured the amount of absorption toward each molecular 
clump in the Sgr B complex. 
Seven different spectral components (with an even 
larger number of free parameters) contribute to the X-ray spectrum, 
and thus an exceptionally high X-ray spectral quality is required in order to 
constrain the various components without incurring degeneracies 
between the different parameters. Moreover, intensity fluctuations of 
the diffuse soft X-ray plasma (which is known to have a patchy distribution) 
is likely another source of uncertainty. 
The distribution of stars in the central $\sim1$ degree is much more 
uniform. Thus stellar counts in near-IR images can provide 
reliable constraints on the line of sight distance toward molecular 
clumps (see, e.g., Gusten \& Downes 1980; Glass et al. 1987). 
Sawada et al. (2004) proposed a different method based on the 
quantitative comparison between the 2.6-mm CO emission lines 
and 18-cm OH absorption lines. The GC is an intense and presumably 
axisymmetric source of diffuse, non-thermal emission at 18-cm,  
so OH absorption arises preferentially from clouds located in the 
foreground of the GC. On the other hand, CO emission samples the gas 
both in front and back of the continuum source equally (Sawada et al. 2004), 
thus (assuming that the lines are optically thin and that the physical and 
chemical properties of GC clouds are uniform) the OH/CO ratio for 
any given cloud provides information about its line-of-sight distance.

The use of the angular dependence of the EW$_{\rm Fe}$ on the angle 
$\theta$ between the direction of the incident radiation and scattered 
continuum (see \S \ref{refl}) may eventually provide another means of 
estimating the line-of-sight distances to the Fe K$\alpha$ emitting MC
(see Capelli et al. 2012). 
However, a maximum variation of the EW$_{\rm Fe}$ of $\sim50$ \% is 
expected, thus very solid estimates of MC abundances (producing 
EW variations of the order of several; see \S \ref{refl}) and $N_H$ 
(producing EW variations of at least $\sim30$ \%; Matt et al. 2003) 
are required. 

Because of physical interactions between the different molecular filaments 
and streamers, it is easier to estimate the MC distribution in the immediate 
vicinity of Sgr A* (within the central 15 pc). Coil et al. (2000), in fact, 
suggested that the 20 and 50 km s$^{-1}$ MC are located at less than 
15 pc from Sgr A*, just outside the CND, with the former in front 
and the latter a few pc behind Sgr A* (see also Ferriere 2012). 
Perhaps the most promising method for determining the line-of-sight 
distances of some of the MC containing $H_2O$ maser sources is 
based on VLBI parallax measurements. Reid et al. (2009) measured a 
distance from the Earth 
to Sgr B2 of $R_0=7.9^{+0.8}_{-0.7}$ kpc and, assuming a circular orbit 
for Sgr B2, they estimated Sgr B2 to be $\sim130$ pc in front of Sgr A*. 
Although this measurement roughly agrees with the location estimated 
by Ryu et al. (2009), Molinari et al. (2011) suggested an elliptical orbit for 
Sgr B2 which would be then located {\it behind} Sgr A*. 
At the moment, it seems that the line of sight distance of the molecular 
clumps is the major source of uncertainty in measuring the GC's past 
activity (influencing both the derived luminosity and time evolution of 
the event). 

{\bf Column density.} 
The other major source of uncertainty is in the determination of the cloud 
column density. Three main ways to determine the MC column 
densities ($N_H$) have been used: 
1) The $N_H$ can be derived directly from high quality X-ray spectra 
of each molecular clump. In fact, as long as the column density is lower 
than the Compton thick limit ($N_H\lsimeq\sigma^{-1}_T=1.5\times 
10^{24}$ cm$^{-2}$) the MC absorption will leave its signature through 
a clear low energy cut off in the 2-10 keV X-ray spectrum (Comastri 2004). 
However, these measurements trace only the total $N_H$ along the line 
of sight, thus including foreground absorption plus the 
additional contributions from any MC along the line of sight that are not related 
to the Fe K$\alpha$ emitting cloud. Moreover, such measurements are 
affected by the presence 
of many additional spectral components (e.g. 6.7 keV emission, soft X-ray 
plasma, weak point sources) contributing to the observed MC X-ray emission. 
However, a strong improvement is expected soon thanks to hard X-ray 
focussing telescopes (such as \nustar\ and \astroh) that will enhance, by 
more than an order of magnitude, the detector sensitivity in the 10 to 80 
keV energy band. In fact, in optically thin clouds ($\tau\ll1$), 
the shape and intensity of the high energy reflection spectrum depends 
strongly (see the right panel of Fig. \ref{Refl}) on the cloud column 
density (particularly the Compton hump, see \S \ref{Chump}). 
Combining the information from the low energy reflection cut off, 
the line intensity and the Compton hump shape, \nustar\ and \astroh\ 
X-ray spectra will soon 
provide a more reliable $N_H$ estimate. 
2) Alternatively, the $N_H$ can be measured through the intensity of molecular 
emission lines. 
Tsuboi et al. (1999) provided a formulism (assuming a CS to $N_H$ abundance 
$X(CS)=10^{-8}$; Irvine et al. 1987) to derive the cloud $N_H$ from 
the CS line intensity (I$_{CS}$): 
$N_H=\frac{7.5\times10^{11}\times T_{ex}\times I_{CS}}{X(CS)}$, 
where $T_{ex}$ is the excitation temperature. 
The great advantage of this method is that it allows us to measure the $N_H$ 
associated with each single molecular clump. In fact, any intervening material is expected
to move at a different velocity, thus the extremely high spectral resolution 
with which it is possible to measure MC lines allows us to slice the CMZ in different 
velocity components, and to select MC of interest. However, large uncertainties 
are associated with $T_{ex}$ and the $X(CS)$ factor, unless detailed modelling 
is performed. Moreover, at high column densities ($N_H>$several$\times10^{23}$ 
cm$^{-2}$) most molecular lines become self-absorbed and hence are no 
longer good $N_H$ estimators. 
3) Molinari et al. (2010) measured the cloud $N_H$ using IR \herschel\ 
observations. They first estimated the opacity and dust temperature from a 
pixel-to-pixel fit on the $70-350 \mu$m data, using a dust model (Compiegne 
et al. 2011; Bernard et al. 2010). Then, assuming constant 
dust properties, they converted the opacity to a hydrogen column density 
($N_H=N_{HI}+N_{H2}$) using $\tau_{250}/N_H=8.8\times10^{-26}$ 
cm$^2/H$. The gas to dust ratio used is valid for the Galactic disc. 
Uncertainties of a factor of a few are associated with these measurements. 
In particular, different metallicities with respect to the disc values, would 
lower the estimated column densities by a factor of a few.
Moreover, as with the $N_H$ estimated from X-ray 
absorption, the column densities derived with this method integrate the 
effects of dust along the line of sight. 

\subsection{Sgr A*'s X-ray emission history} 
\label{SgrA*history}

The study of the Fe K$\alpha$ emission from the MC of the CMZ shows 
strong indications of external irradiation, most probably due to Sgr A*. 
We also note that: i) spatially connected structures (such as Sgr B1 
and B2 or the bridge) tend to vary in brightness with a coherent pattern; 
ii) molecular complexes that appear to be located close together, but display 
different Fe K$\alpha$ light curves (such as the bridge, MC1, MC2 and 
G0.11-0.11; Ponti et al. 2010), belong to molecular complexes moving
at different line-of-sight velocities. 
The latter point indicates that these molecular complexes are actually 
well separated, in three dimensions, despite being close together 
on the plane of the sky. 
The different light curves observed from these complexes could then 
be tracing either different flares or different parts of the same long flare. 
Moreover, we note that smaller MC generally vary on faster time-scales 
(with the variation fronts usually propagating close to, or higher than, the 
speed of light) compared to larger MC. The latter varying on a time scale
being typically of the order of the cloud light crossing time. 
This suggests that the Fe K$\alpha$ emission is induced by irradiation 
by an external source that is variable on both long and short time-scales. 
Such variability behaviour is usually observed in, for example, sources 
powered by accretion onto compact objects (Belloni \& Hasinger 1990; 
McHardy et al. 2006; Koerding et al. 2007; Ponti et al. 2012).

Constraints on the GC's past activity are given by the existence of MC that 
emit Fe K$\alpha$ and by those clouds that are quiet. Only upper limits to the 
Fe K$\alpha$ emission from the MC closest to Sgr A* (the CND, the 20 
and 50 km s$^{-1}$ MC; Coil et al. 2000) are observed. If, indeed, the 50 km 
s$^{-1}$ MC is located 5-10 pc behind Sgr A* (Coil et al. 2000), then from the 
estimated column density and the upper limit on the line intensity, we can 
constrain the mean luminosity of Sgr A* to be lower than 
$L_{\rm Sgr A*} \lsimeq 8 \times 10^{35}$ erg s$^{-1}$ in the past 60-90 yr
(Ponti et al. 2010). 
The observation of approximately constant Fe K$\alpha$ emission during 
the past decade, from molecular complexes with sizes smaller than a parsec, 
indicates that the flare illuminating those clouds lasted more than several 
years. Moreover, the observation of both a rising and fading phase from some 
molecular complexes (Ponti et al. 2010; Capelli et al. 2011) suggests, not surprisingly, 
some degree of variability of the illuminating source. In fact, the light curves of 
accretion powered sources are well known to be variable 
(Belloni \& Hasinger 1990; McHardy et al. 2006; Koerding et al. 2007; Ponti et al. 2012).

The uncertainty in the line of sight distance toward the different molecular 
clouds in the CMZ currently prevents us from putting any firm 
constraint on the flare history, apart from knowing that Sgr A* underwent 
one or more flares with luminosities of the order of $10^{39}$ erg s$^{-1}$ 
or higher, between $\sim100$ and $\sim1000$ years ago (and possibly fainter 
ones). Assuming that Sgr B2 is located $\sim130$ pc in front of Sgr A*, then 
the flare must have ended $\sim100$ yr ago. 
The Fe K$\alpha$ evolution of G0.11-0.11 suggests that this cloud is 
illuminated by the same phase of the flare currently illuminating Sgr B2. Further 
observations will allow this hypothesis to be tested. 
Actually, if there is just a single (or a handful) illuminating source, we expect that 
eventually every MC will reflect the same light curve (convolved with 
the MC dimension). Therefore, it will eventually be possible to measure 
the delay of the light curve and hence determine the line of sight distance
for each MC. Alternatively, if the line-of-sight to the clouds is known, it will 
be possible to triangulate the position of the primary source. 
The light echos provide us with the fantastic opportunity to 
scan the CMZ and determine its three dimensional distribution. 

Although a precise number has not yet been measured, about one 
third of the MC in the CMZ are Fe K$\alpha$ active. This suggests 
that in the past $\sim10^3$ yr (the CMZ light crossing time), Sgr A* 
was active for a significant fraction of the time. Longer monitoring will 
clarify if the higher activity was made of several sharp excursions 
from quiescence to a short lived bright state, or if it involved a more gentle 
evolving and a longer period of activity. The constraints are still so weak 
that a single long period of activity is still consistent with the observations 
(Ponti et al. 2010, but see Capelli et al. 2012). 
New observing campaigns will provide stronger constraints, 
allowing us to learn much more about Sgr A*'s past activity. 

Is it possible to constrain Sgr A*'s luminosity further in the past?
To answer this question, Cramphorn \& Sunyaev (2002)
studied the Fe K$\alpha$ emission from all the MC in the Galactic disc.
None of the MC they observed showed any evidence of Fe K$\alpha$
emission. This allowed the authors to put several upper limits on Sgr A*'s 
historical luminosity. Sgr A*'s average luminosity 
was found to be lower than $\sim10^{-(3-4)}$ L$_{\rm Edd}$ 
since $\sim4\times10^4$ yr ago. Obviously the disc of the Milky Way is 
not completely filled with molecular gas, so gaps in the CO distributions 
leave unconstrained periods as long as $2-4\times10^3$ yr.
To extend this study even further, Cramphorn \& Sunyaev (2002) 
investigated also the more extended HI distribution, allowing them 
to constrain Sgr A*'s past activity to L$_{Sgr A*}<10^{-2}$ L$_{\rm Edd}$, 
in the last $\sim1\times10^5$ yr. 

To extrapolate even further into the past, different types of tracers of 
Sgr A*'s luminosity are necessary. 
The presence of the \fermi\ bubble, the EMR and the disc of young 
stars within the central parsec, suggest that Sgr A* was radiating close to 
the Eddington limit $\sim5-6\times10^6$ years ago.

\subsection{Is all the Fe K$\alpha$ emission due to Sgr A*'s flares?}

The measurements of fast Fe K$\alpha$ flux variations (Inui et al. 2009; 
Terrier et al. 2010; Nobukawa et al. 2011; Ponti et al. 2010; Capelli et al. 2011; 
2012) and even the super-luminal propagation of the X-ray echo (Ponti et al. 2010) 
have excluded a cosmic ray proton/ion origin for this 
variable component (Fe K$\alpha$ emission induced by cosmic ray electrons 
might vary on years timescales; Yusef-Zadeh et al. 2012) and cosmic ray 
electron and internal source for the 
super-luminal one. This implies that at least part of the Fe K$\alpha$ 
radiation has to be produced by a bright external source, most probably 
Sgr A* (see \S \ref{whichsource}). However, is it plausible to think that all 
Fe K$\alpha$ radiation from MC of the CMZ is induced by Sgr A*?

Clear signatures for cosmic ray ion induced Fe K$\alpha$ emission 
have been discovered in molecular gas around the Arches cluster 
(Wang et al. 2006; Tsujimoto et al. 2007; Capelli et al. 2011; 
Tatischeff et al. 2012). The prominent (EW$\sim1.2$ keV) Fe K$\alpha$ 
emission has, in fact, been observed to be constant over the past 
8 yr (Capelli et al. 2011). Moreover the Fe K$\alpha$ emission is not well 
correlated with the molecular gas, forming, instead, a bow shock around 
the cluster (Wang et al. 2006). 
The Arches cluster is one of the richest and most densely 
packed massive ($M\sim5\times10^4$ M$_{\odot}$; Harfst et al. 2010) 
star clusters in the Milky Way and is currently moving at 
highly supersonic speed ($v_{\rm Arch}\sim200$ km s$^{-1}$) relative to 
the bulge stars. Figer et al. (2002) and Wang et al. (2006) found 
that the cluster bow shock is interacting with a MC and Tatischeff et 
al (2012) computed that the shock (assuming a reasonable particle 
acceleration efficiency of $\sim3-10$ \%) could supply enough cosmic 
ray power to reproduce 
the Fe K$\alpha$ emission. Therefore not every Fe K$\alpha$ active 
cloud is illuminated by a flare from Sgr A*. 

A deep \suzaku\ observation of the molecular cloud G0.162-0.217
revealed a weak Fe K$\alpha$ line with EW$_{\rm FeK\alpha}\sim0.2$ 
keV (Fukuoka et al. 2009). The cloud is located at the south end of the 
Radio Arc (Yusef-Zadeh et al. 1984; LaRosa et al. 2000; Yusef-Zadeh 
et al. 2004), a region with, supposedly, enhanced density of cosmic ray 
electrons. Fukuoka et al. (2009) thus proposed that this emission might be 
induced by cosmic ray electrons. 

\section{Origin of the flares-outbursts}

The study of the environment around Sgr A* suggests at least two periods of 
enhanced activity for the supermassive BH at the Milky Way centre 
(see \S \ref{SgrA*history}), the first about $6\pm2\times10^{6}$ yr ago, 
with Sgr A* emitting close to the 
Eddington limit, and the second with a luminosity $1-5\times10^{39}$ erg 
s$^{-1}$, which occurred $\sim10^2$ yr ago. 

Traces of what triggered the earlier of these accretion events can 
be found in the close proximity of Sgr A*. The cluster of young stars in the 
central parsec (see \S \ref{youngstars}) suggests the presence of a very 
massive disc there, $\sim6\times10^{6}$ yr ago. 
Some have suggested (although the origin of such disc is still controversial)
that this disc might have formed from a MC passing very close to Sgr A*. 
As a result of tidal disruption and stretching, the cloud would follow a range of 
intersecting orbits, near the circularisation radius, forming an eccentric, 
clumpy and filamentary disk. The physical conditions in the disk would 
allow star formation and accretion onto the BH (Sanders 1998; 
Bonnell \& Rice 2008; Genzel et al. 2010; Yusef-Zadeh \& Wardle 2012). 
Part of the matter in the inner part of this disc would have been accreted, 
making Sgr A* shine close to the Eddington limit. Part of the disc would have 
formed massive stars (the central cluster). The outer part of the disc might have 
been left as a remnant disc and could now comprise what we call the CND 
(see \S \ref{CND}). Finally, a part of the disk would have formed an outflow, such 
as commonly seen in luminous AGN, perhaps generating both a jet 
(Su \& Finkbeiner 2012) and inflating the \fermi\ bubbles (Su et al. 2010). 

Somewhat more ambiguous is the physical origin of the enhanced emission 
that occurred $\sim10^2$ yr ago. Sgr A*'s quiescent X-ray luminosity 
(see \S \ref{SgrA*}) is $L_{\rm 2-10~keV}=2\times10^{33}$ erg s$^{-1}$ 
(Baganoff et al. 2003) and even during the brightest X-ray flare ever observed 
(in more than 10 years of \xmm\ and \chandra\ monitoring) the X-ray luminosity 
was only $L_{\rm 2-10~keV}=3\times10^{35}$ erg s$^{-1}$ (Porquet et al. 2003). 
The luminosity required to produce the X-ray echo seen from the MC, is 
$L_{\rm 2-100~keV}=1.5-5\times10^{39}$ erg s$^{-1}$ 
(Terrier et al. 2010) and is about $\sim10^4$ times higher than the brightest flare 
ever observed. This seems to suggest an intrinsic difference between 
the process producing the daily flaring activity and this major flare-outburst.
It is rather surprising then that the required X-ray luminosity can be inferred 
by simply extrapolating the observed K-band flux distribution, obtained over 
$\sim10$ years of monitoring, to higher fluxes and rare events (Witzel et al. 2012, 
Dodds-Eden et al. 2011). 
This extrapolation to high fluxes and the assumed IR to X-ray conversion involve 
many assumptions (such as the flare SED and a direct correspondence 
between the IR and X-ray light curves). Despite these uncertainties, the extrapolation 
seems to suggest that the X-ray echo is simply an extreme manifestation 
of the same physical processes that produce the daily flux variability. 
From a theoretical point of view this conjecture seems 
plausible. In fact, during normal flux states Sgr A* is thought to accrete from 
the capture of stellar wind material from nearby stars (see \S \ref{SgrA*}). 
Cuadra et al. (2008) suggested that, during this process, cold clumps might 
sometimes fall into the inner region with an angular momentum low enough 
to circularise at a very small radius ($\sim$0.001 arcsec), and create there 
an accretion disc. Such an event might have enhanced the luminosity to 
$L_{Bol}\sim10^{40}$ erg s$^{-1}$ for a timescale of the order of 
$\sim10^2$ yr (dominated by the disc viscous timescale; Cuadra et al. 2008)
and have produced the X-ray luminosity required for the observed 
Fe K$\alpha$ echo.

Yu et al. (2011) suggested, instead, that Sgr A*'s flare could be produced 
by the shock created by the jet deceleration (and not by accretion) caused 
by a partial capture of a star by the supermassive BH. This model well 
reproduces the Fe K$\alpha$ emission (both the flux and its evolution) 
of Sgr B2, with a decay time of $\sim10$ yr (Nobukawa et al. 2011;
Terrier et al. 2010; Inui et al. 2009; Revnivtsev et al. 2004). 
Zubovas et al. (2012) consider the consequences of an asteroid  
passing within a few astronomical units from Sgr A* and being gravitationally 
disrupted. Objects with radius 
$r\sim10-50$ km are expected to produce $L_x\lsimeq10^{36}$ erg s$^{-1}$,
thus possibly contributing to Sgr A*'s normal daily flaring activity. However, 
a planet disruption event (although expected to be much less frequent, 
e.g. one per $\sim10^5$ yr)
might produce flares with $L\sim10^{41}$ erg s$^{-1}$ for a few tens of years 
(Zubovas et al. 2012).
Alternatively, Sazonov et al. (2012) considered gas supply onto Sgr A* via tidal 
interaction of stars. They estimated that every few $10^4$ yr, less than 
$0.1$ M$_{\odot}$ of stellar debris will accrete onto Sgr A* and produce 
an outburst of $L\sim10^{42-43}$ erg s$^{-1}$. A lower luminosity flare 
could be produced by gas torn off a star that experienced a close collision 
with another star or a stellar remnant in the nuclear cluster. 

Accretion of interstellar material is perhaps a more probable and frequent 
occurrence because of its larger cross section. In fact, the vicinity of Sgr A*
contains several ionised gas streams (Zhao et al. 2009) and numerous 
blobs of dust and gas (Ghez et al. 2005; Muzi\'c et al. 2010). 
Thanks to the analysis of more than ten years of infra-red monitoring observations
of the stellar orbits of the stars close to Sgr A*, it has been recently discovered that 
a dense cloud (G2) with a mass of at least $\sim$$3$ M$_{\rm Earth}$ (estimated 
from the Br$\gamma$ luminosity), is now moving (and accelerating) towards Sgr A* 
(Gillessen et al. 2012a,b). The centre of gravity of the cloud should be at the 
pericentre (at only 3100 times the BH event horizon) in mid-2013. 
Due to the interaction between the gas in the cloud and the ambient 
material, dynamical instabilities (such as Rayleigh-Taylor and Kelvin-Helmholtz 
instabilities) will disrupt and fragment the cloud into pieces, some of which might 
accrete onto Sgr A*. It is predicted that the cloud will generate a small 
increase in the 2-8 keV luminosity of only one order of magnitude above the 
quiescent level (Burkert et al. 2012; Schartmann et al. 2012). 
However, the cloud origin, mass content and the effects for the future activity 
of Sgr A* are still highly debated (Miranda-Escude 2012; Narayan et al. 2012; 
Anninos et al. 2012). Murray-Clay \& Loeb (2012) suggested that this cloud 
naturally originates from a proto-planetary disc surrounding an undetected 
low mass star. If so, the mass accreted onto Sgr A* (and thus Sgr A*'s
luminosity) might be orders of magnitude higher producing a rare and 
bright flare. Further observations of G2 will give us an unprecedented 
opportunity to study one possible mode of accretion onto a supermassive 
BH.

\section{An AGN torus in our backyard!}

Several pieces of evidence suggest that Sgr A* might have undergone 
a period of AGN activity about $6\times10^6$ yr ago (see \S \ref{SgrA*history}; 
Zubovas et al. 2011; 2012). 
In those conditions the CMZ would have appeared, to an extragalactic 
observer, to have the properties of a typical AGN torus commonly invoked by AGN 
unification models (Antonucci 1993; Urry \& Padovani 1995). 
In fact, it would have obscured (with column densities up to 
$N_H\sim10^{24-25}$ cm$^{-2}$) the direct vision of the central BH to 
an equatorial observer. At about 50-150 pc the height of the CMZ is about 
several tens of parsec, thus generating obscuration toward a significant fraction of 
line of sights. This number is not too dissimilar from the one typically 
estimated from the fraction of obscured/unobscured AGN (Urry \& Padovani 1995;
Risaliti et al. 1999). 
The kinematics of the EMR (see \S \ref{EMR}) is modified by the influence of 
the stellar bar potential (Binney 1994). However, the radial component of the 
gas motion could have been generated by the radiation pressure and/or the 
momentum of an outflow produced by accretion onto Sgr A*, that might have 
pushed away the molecular material from the accretion disc axis 
(Sofue et al. 1995a,b), carving two holes of lower extinction. 
However, along the Galactic disc, where most of the molecular mass is 
concentrated, the CMZ represents an almost impassable barrier to Sgr A*'s 
radiation/outflows. 
The mass of the CMZ is in fact large enough to impede equatorial outflows 
from Sgr A*, even if accreting at the Eddington limit 
($L_{\rm Edd}\sim5\times10^{44}$ erg s$^{-1}$). 
During the last AGN phase, the inner faces of the inner CMZ clouds would 
have seen the nuclear continuum 
(efficiently producing UV and X-ray radiation; Elvis et al. 1994)
unobscured and be heated by photoionisation, initially to temperatures 
of $\sim10^4$ K, then to higher temperatures as their material evaporates
away (Krolik \& Begelman 1988). Part of this gas might have generated an outflowing 
wind (Krolik \& Kriss 1995; 2001), possibly forming a warm absorber 
component such as observed in UV/X-ray spectra of AGN 
(Blustin et al. 2005; Kaastra et al. 2012; Kriss et al. 2011), 
while the remaining matter would have probably been channelled inwards 
at a rate sufficient to maintain a steady state against evaporation of the torus inner edge 
(provided that new material can be fed into the outer edge of the torus; 
Krolik \& Begelman 1988). The luminosity of the nuclear source sets a limit to the 
survival of dust. Therefore, the so called "dust sublimation radius" is typically 
assumed to be the inner edge of the molecular torus (Barvainis 1987). 

The advent of mid and near-IR interferometric observations at VLT and Keck 
has recently allowed a dramatic improvement in angular resolution. The application 
of this technique to a few nearby AGN has allowed investigators to marginally and 
partially resolve the dust sublimation region (the inner edge of the torus) in these objects 
(Swain et al. 2003; Jaffe et al. 2004; Tristam et al. 2007; Meisenheimer et al. 2007;
Burtscher et al. 2009; Kishimoto et al. 2011a,b). 
Near-IR observations of AGN (tracing hot, $T\sim10^3$ K, dust emission) reveal 
geometrically-thin ring radii on sub-parsec ($\sim0.03-0.5$ pc) scales, 
consistent with the near-IR reverberation radius (the light crossing distance 
corresponding to the delay between the variabilities in the UV/optical and near-IR; 
Kishimoto et al. 2011). Moreover, the ring radii present a dependence 
with source luminosity, as expected for the dust sublimation radii (Kishimoto et 
al. 2011a,b; Barvainis 1987). 
Mid-IR observations (tracing warm, $T\sim10^{2-3}$ K, 
dust) display significantly larger radii with a size going from a fraction up to 
several parsecs. This clearly indicates a stratification in temperature of the radial 
structure of the torus, with the hot near-IR emission tracing the inner torus edge, 
closer to the irradiating source. 
IR observations are thus dominated by a central brightness concentration 
at the scale of the sublimation radius, the bright rim of the innermost dust 
distribution, and weaker emission from the external, weakly irradiated, part 
of the torus (Kishimoto et al. 2011a,b). More difficult is to determine the 
structure of the outer, weaker emission or any substructure within these AGN tori. 

These results do not contradict the idea that the CND might be the 
relic of the inner part of the Milky Way AGN torus. Once the intense feeding 
onto Sgr A* stopped, photo-evaporation halted and the CND has been left as a relic
(Duschl 1989) just outside the sublimation radius. The orbital time of the CND material 
is much shorter than $10^6$ yr, so it is unlikely that it kept the same shape 
that it might have had during the AGN phase. 
After a long period of quiescence, we do not expect to observe remnant 
radiation from the emitting regions surrounding a typical AGN, i.e.  
the broad and narrow line regions, the warm absorber, etc. In fact, 
the recombination timescales of such plasmas are expected to be much 
shorter than $\sim10^6$ yr (Osterbrock \& Mathews 1986). 

Being only $\sim8$ kpc away (i.e. $\sim500$ times closer than the 
closest AGN, Cen A), the GC provides us with the unique opportunity to 
resolve spatially the internal structure of an AGN torus in such exquisite detail. 
It is expected that the outer CMZ structure has not changed significantly since 
the last AGN phase ($\sim6\times10^6$ yr ago). 
In fact, a typical MC core, at $\sim100$ pc from the centre, moving 
at $\sim100$ km s$^{-1}$ (typical orbital velocities of a MC of the CMZ disc 
population) would require about $10^7$ yr to complete 
a full orbit. This suggests that the external structure of the past AGN torus 
is still preserved in the CMZ. Thus the CMZ can be used as a template 
to study the structure of the tori of distant AGN and the mass accretion 
from the outer to inner boundary of AGN tori.  

The study of Sgr A*'s surroundings has allowed us to hint at its past activity and 
the impact of Sgr A*'s emissions on the surrounding matter and star forming 
regions. Zubovas et al. (2011) estimate that, to power the last AGN phase, 
Sgr A* might have accreted about $\sim2\times10^3$ M$_{\odot}$. This is 
about one order of magnitude smaller than the estimated mass stored in the CND. 
The irregular and clumpy structures of the CND, suggest dynamical evolution 
and episodic feeding of gas towards Sgr A* (Liu et al. 2012). The CND appears
to be the convergence of the innermost parts of large-scale gas streamers, 
which are responding to the central gravitational potential well. 
All these indications suggest that the CND might feed the supermassive BH again. 
Morris et al. (1999) proposed a limit cycle of recurrent nuclear activity 
where, in the absence of dynamical or radiation pressure from within 
the CND cavity, viscous disc evolution would cause the inner edge of the CND 
to shrink, eventually fuelling the BH. The associated burst of radiation would 
presumably produce shocks and compression of the surrounding gas,  
possibly inducing a dramatic episode of star formation that could produce 
a massive cluster like the present central parsec cluster of young stars. 
The collective winds and radiation pressure from the newly-formed stars 
and/or AGN accretion, might clear gas from the central region, halting 
accretion and creating an inner cavity in the CND. 
The stellar winds from the massive cluster of newborn stars fade out 
about $\sim10^7$ yr after the star formation event and, at that point, the inner 
edge of the CND would start migrate inward again, thus completing the 
cycle (Morris et al. 1999).

On larger scales, studies of the CMZ structure also suggest that episodic cloud 
formation and starbursts in the GC region (with a relaxation oscillator mechanism 
for quasi-periodic starbursts of $\sim10^{6}-10^{7}$ years; Stark et al. 2004; 
Martin et al. 2004; Morris \& Serabyn 1996) are a natural result of the interaction 
between the stellar bar and interstellar gas. 
This star formation cycle, as well as the cloud kinematics and evolution, 
can be influenced by Sgr A*'s activity, which in turn will be triggered by matter 
migrating into its sphere of influence. 
Therefore, the study of Sgr A*'s past activity and its connection to observable 
phenomena in the CMZ provide a high-resolution opportunity for studying 
the AGN feedback phenomenon.

\begin{acknowledgement}
The authors thank the ISSI in Bern. GP thanks S. Bianchi, E. Churazov, 
G. Matt, S. Gillessen, V. Tatischeff, T. Dwelly, A. Strong, L. Burtscher, 
A. Merloni, Y. Tanaka and F. Melia for useful discussion. 
Part of the work was supported by the French Labex "UnivEarthS".
\end{acknowledgement}

\end{document}